\documentclass[prd,reprint,nofootinbib,superscriptaddress]{revtex4-2}
\usepackage{bm}
\usepackage{amsmath}
\usepackage{amssymb}
\usepackage{graphicx}
\usepackage{float}
\usepackage{subfigure}
\usepackage{hyperref}
\usepackage{CJK}
\hypersetup{
colorlinks=true,
linkcolor=red,
citecolor=blue,
}

\begin{document}

\title{Parameter estimation for space-based gravitational wave detectors with ringdown signals}

\author{Chunyu Zhang}
\email{chunyuzhang@hust.edu.cn}
\affiliation{School of Physics, Huazhong University of Science and Technology,
Wuhan, Hubei 430074, China}

\author{Yungui Gong}
\email{Corresponding author. yggong@hust.edu.cn}
\affiliation{School of Physics, Huazhong University of Science and Technology, Wuhan, Hubei 430074, China}

\author{Chao Zhang}
\email{chao\_zhang@hust.edu.cn}
\affiliation{School of Physics, Huazhong University of Science and Technology,
Wuhan, Hubei 430074, China}

\begin{abstract}
Unlike ground-based gravitational wave detectors, space-based gravitational wave detectors can detect the ringdown signals from massive black hole mergers with large signal-to-noise ratios, help to localize sources and extract their parameters.
To reduce the computation time in the Fisher information matrix analysis, we derive the analytical formulas of frequency-domain ringdown signals for both heliocentric and geocentric detectors by considering the effects of the harmonic phases, the rotation period of the geocentric detector, and the detector's arm length.
We explore median errors of the parameter estimation and source localization with ringdown signals from binaries with different masses and different redshifts.
Using a binary source with the total mass $M=10^7\ M_\odot$ at the redshift $z=1$, 
we analyze the dependence of these errors on the sky position.
We find that the network of space-based gravitational wave detectors can significantly improve the source localization at the ringdown stage.
The results of the Fisher matrix approximation are also checked by Bayesian inference method.
\end{abstract}


\maketitle

\section{Introduction}

The detection of gravitational waves (GWs) by the Laser Interferometer Gravitational-Wave Observatory (LIGO) Scientific Collaboration and the Virgo
Collaboration not only announced the dawn of a new era of multimessenger astronomy, but also opened a new window to probe the nature of gravity and spacetime in the nonlinear and strong field regimes \cite{Abbott:2016blz,Abbott:2016nmj,TheLIGOScientific:2016agk,Abbott:2017vtc,Abbott:2017oio,TheLIGOScientific:2017qsa,Abbott:2017gyy,LIGOScientific:2018mvr,Abbott:2020tfl,Abbott:2020khf,LIGOScientific:2020stg,Abbott:2020uma,Abbott:2020niy}.
GWs from compact binary coalescences consist of inspiral, merger, and ringdown phases, with increasing frequency.
The inspiral waves, at the stage of orbiting until the innermost stable orbit, can be analyzed by the post-Newtonian theory, black hole (BH) perturbation theory, etc.
At the early inspiral stage, the emitted GWs can be regarded as monochromatic waves due to the slow orbital decay.
The merger waveform which is not well modeled at present, 
is the research topic in numerical relativity.
The ringdown signal originating from the distorted final BH, 
comprises a superposition of quasinormal modes (QNMs).
The frequency of each mode is a complex number, 
the real part is the oscillation frequency, 
and the imaginary part is the inverse of the damping time.
These frequencies are determined by the mass $M$ and angular momentum $J$ of the final BH, 
and the amplitude and phase of each mode are determined by the specific process when the final BH forms.

Ground-based GW detectors, such as Advanced
LIGO \cite{Harry:2010zz,TheLIGOScientific:2014jea}, 
Advanced Virgo \cite{TheVirgo:2014hva} and Kamioka Gravitational Wave Detector (KAGRA) \cite{Somiya:2011np,Aso:2013eba}, 
operate in the $10-10^4$ Hz frequency band.
In this frequency band, 
the detected events are stellar-mass binary mergers, 
the ringdown signal is not loud enough to probe the physics behind it.
The proposed space-based GW detectors such as Laser Interferometer Space Antenna (LISA)
\cite{Danzmann:1997hm,Audley:2017drz}, TianQin \cite{Luo:2015ght}, and Taiji \cite{Hu:2017mde} probe GWs in the millihertz frequency band, while Deci-hertz Interferometer Gravitational Wave Observatory (DECIGO) \cite{Kawamura:2011zz} operates
in the 0.1 to 10 Hz frequency band.
Thus, space-based GW detectors can detect ringdown signals from massive BH binary mergers with large signal-to-noise ratios (SNRs), 
and the detected  ringdown signals can be used to probe the nature of BHs,
localize sources and estimate their parameters, etc.
In particular, the sky localization of the source is one of the important scientific objectives for GW observations
because accurate information about the source localization is necessary for the follow-up observations of electromagnetic counterparts and the statistical identification of the host galaxy if no counterpart is present. 
Therefore, cosmological applications such as studying the problem of Hubble tension \cite{Riess:2019cxk}
using GWs as standard sirens \cite{Schutz:1986gp, Holz:2005df} depend critically on the capability of locating the source accurately.

In general, higher multipoles and higher harmonics are subdominant in the inspiral phase,
so usually the parameter estimation for space-based GW detectors was analyzed with Fisher information matrix (FIM) method by considering the (2, 2) mode only \cite{Peterseim:1996cw,Peterseim:1997ic,Cutler:1997ta,Cutler:1998muh,Moore:1999zw,Barack:2003fp,Blaut:2011zz,Vallisneri:2007ev,Wen:2010cr,Aasi:2013wya,Grover:2013sha,Berry:2014jja,Singer:2015ema,Becsy:2016ofp,Zhao:2017cbb,Mills:2017urp,Fairhurst:2017mvj,Fujii:2019hdi,Ruan:2019tje,Ruan:2020smc,Feng:2019wgq,Wang:2020vkg,Huang:2020rjf,Zhang:2020hyx,Zhang:2020drf,Shuman:2021ruh,Mangiagli:2020rwz}.
The parameter estimation as function of time left before merger was also discussed in \cite{Mangiagli:2020rwz}.
Higher multipoles and higher harmonics of GWs have characteristic structure in the gravitational waveforms,
and they have different dependence on the source's parameters such as the inclination,
the mass ratio and spins,
so they can be used to break some of the degeneracies between the parameters and improve the parameter estimation accuracy \cite{Barausse:2009xi,Pan:2010hz,Kamaretsos:2011um,Kamaretsos:2012bs,London:2014cma,Baibhav:2017jhs,Baibhav:2018rfk}.
Furthermore, higher harmonics can break the degeneracy between the polarization and the coalescence phase \cite{Payne:2019wmy}.
Because the contribution of higher multipoles to the radiated energy increases with the mass ratio \cite{Buonanno:2006ui,Berti:2007fi},
the estimations of the mass ratio and the effective spin are improved with higher harmonics for the GW170729 event \cite{Chatziioannou:2019dsz}.
For heavier binaries with $M\gtrsim 10^7\ M_\odot$,
higher harmonics of inspiral can improve the angular resolution of LISA by a factor of $\sim 100$ \cite{Arun:2007qv,Arun:2007hu,Trias:2007fp,Porter:2008kn}.
If the total mass of a binary is very large,
then the inspiral waves are out of the frequency band of space-based GW detectors and the ringdown waves with higher harmonics become dominant because they have higher frequencies \cite{Flanagan:1997sx,Rhook:2005pt,Berti:2005ys}.
For ringdown GWs, ignoring the source location,
the capability of TianQin to test the no-hair theorem of general relativity was studied \cite{Shi:2019hqa}.
With higher harmonics of the ringdown signals from binaries with the total mass $M\ge 10^6\ M_\odot$,
employing the error propagation method in FIM and ignoring the influences of the arm length and harmonic phases,
the parameter estimation and source localization of LISA were studied in Ref. \cite{Baibhav:2020tma}.
Compared with Bayesian inference method,
the FIM method provides poor estimations for the extrinsic parameters such as sky location, luminosity distance and inclination angle
and we cannot make strong statements about parameter estimation with massive BH binaries using the FIM method \cite{Porter:2015eha,Cornish:2020vtw,Marsat:2020rtl}.
It was shown that higher harmonics can break degeneracies between parameters and considerably improve the source localization
of massive BH binaries with LISA by using Bayesian parameter estimation \cite{Marsat:2020rtl}.
With Bayesian inference method,
varied correlations between the total masses and mass ratios and the ability of sky localization of the source with LISA were discussed by analyzing seven test  massive BH binaries using the PhenomHM waveform with higher harmonics and aligned spins \cite{Katz:2020hku}.

Note that to perform parameter estimation, we need to identify the presence of a signal first.
In reality, data gaps, glitches, nonstationary and non-Guassian noises, orbital evolution, unequal arms and superposed signals of different types, all increase the complexity of data analysis \cite{Cornish:2020vtw,Marsat:2020rtl,Crowder:2004ca,Memmesheimer:2004cv,Konigsdorffer:2005sc,Cuoco:2004tr,Cornish:2014kda,Carre:2010ra,Tinto:1999yr,Armstrong_1999,Dey:2021dem}.
To cancel the large laser frequency noise in an unequal arm interferometer detector,
Time-delay interferometry was proposed in \cite{Tinto:1999yr,Armstrong_1999}.
Dey {\it et al.} found that the effect of data gaps due to regular maintenance of the spacecraft
on the detection and parameter estimation of massive BH binaries with LISA is negligible
\cite{Dey:2021dem}.
In this paper, we leave aside these problems and focus on the inference of BH parameters.

The purpose of this paper is to derive analytical frequency-domain detector signals at the ringdown stage, and use them to make parameter estimation.
The  paper is organized as follows.
In Sec. \ref{SecMethod}, we give the analytical formulas of frequency-domain ringdown signals for both heliocentric and geocentric detectors by considering the influences of the harmonic phases, 
the rotation period of the geocentric detector, and the detector's arm length.
The integration formulas we used are presented in Appendix \ref{AppAnalytical}.
In Sec. \ref{SecResults}, 
we show the median errors of the parameters and the source localization with ringdown signals from binaries with different total masses and different redshifts.
We also analyze the dependence of these errors on the sky position.
Then we explore the localization capability of different detectors including the network of space-based GW detectors.
In Sec. \ref{SecBayesian}, we choose two binaries for parameter estimation with Bayesian inference method to check the FIM results.
We conclude this paper in Sec. \ref{SecConclusion}.
Throughout this paper we use units in which $G=c=1$.

\section{Methodology}
\label{SecMethod}	

\subsection{Ringdown waves}
Distorted BHs, such as the newly formed remnant after the coalescence of two BHs, are expected to emit characteristic radiation in the form of QNMs, called ringdown waves, with discrete frequencies.
We usually use three indices $(l,m,n)$ to label the QNMs, where $n=0,1,2,...$ is the overtone index, and $l=2,3,4,...$ and $m=0,\pm 1,...,\pm l$ are the harmonic indices.
Compared to higher overtones with $n\ge1$, the fundamental modes with $n=0$ usually have much larger amplitudes and much longer damping times. 
Thus we only consider the fundamental modes with $n=0$ and denote them as $(\ell, m)$.
Due to the similar reason, and to avoid large numerical-relativity errors, we only use the four strongest modes $(\ell,m)=(2,2),(3,3),(2,1),(4,4)$,
\begin{equation}\label{Eqht}
\begin{split}
h_+(t)&=\frac{M_z}{d_L}\sum_{\ell,m}A_{\ell m}Y^{\ell m}_+(\iota) e^{-\frac{t}{\tau_{\ell m}}}\cos(\omega_{\ell m}t-\phi_{\ell m}),\\
h_\times(t)&=-\frac{M_z}{d_L}\sum_{\ell,m}A_{\ell m}Y^{\ell m}_\times(\iota) e^{-\frac{t}{\tau_{\ell m}}}\sin(\omega_{\ell m}t-\phi_{\ell m})
\end{split}
\end{equation}
for $t>t_0$; and $h_{+,\times}(t)=0$ for $t<t_0$.
Here $t_0$ is the start time of the ringdown waves, $M_z$ is the redshifted mass of the remnant, $d_L$ is the luminosity distance to the source, $A_{\ell m},\omega_{\ell m},\tau_{\ell m}$ and $\phi_{\ell m}$ are the amplitude, oscillation frequency, damping time, and initial phase of the corresponding QNM respectively, and $\iota\in[0,\pi]$ is the inclination angle of the source.

The functions $Y^{\ell m}_{+,\times}(\iota)$ corresponding to the two ringdown polarizations can be found by summing over modes with positive and negative $m$:
\begin{equation}
\begin{split}
Y^{\ell m}_+(\iota)  \equiv & {_{-2}Y^{\ell m}}(\iota,0) + (-1)^\ell\,  {_{-2}Y^{\ell -m}}(\iota,0),\\
Y^{\ell m}_\times(\iota)  \equiv & {_{-2}Y^{\ell m}}(\iota,0) - (-1)^\ell\,  {_{-2}Y^{\ell -m}}(\iota,0).
\end{split}
\end{equation}
For example,
\begin{equation}\label{EqYlm}
\begin{split}
Y^{22}_+(\iota)=&\frac{1}{2} \sqrt{\frac{5}{\pi }} \frac{1+(\cos\iota)^2}{2}, \\
Y^{22}_\times(\iota)=&\frac{1}{2} \sqrt{\frac{5}{\pi }} \cos\iota,\\
Y_+^{21}(\iota) = & \sqrt{\frac{5}{4\pi}}\sin\iota, \\
Y_\times^{21}(\iota) =& \sqrt{\frac{5}{4\pi}} \cos\iota \sin\iota,\\
Y_+^{33}(\iota) = & -\sqrt{\frac{21}{8\pi}} 
\frac{\left ( 1+\cos^2\iota \right )}{2} \sin\iota, \\
Y_\times^{33}(\iota) =& -\sqrt{\frac{21}{8\pi}}\cos\iota \sin\iota,\\
Y_+^{44}(\iota) =& \sqrt{\frac{63}{16\pi}}
\frac{\left ( 1+\cos^2\iota \right )}{2}\sin^2\iota, \\
Y_\times^{44}(\iota) =& \sqrt{\frac{63}{16\pi}} \cos\iota \sin^2\iota.
\end{split}
\end{equation}
The fitting formulas of $\omega_{\ell m}$ and $\tau_{\ell m}$ are given as \cite{Berti:2005ys}

\begin{equation}
\begin{split}
\label{EqOmegaTau}
\omega_{\ell m}&=\frac{f_1+f_2(1-j)^{f_3}}{M_z},\\
\tau_{\ell m}&=\frac{2(q_1+q_2(1-j)^{q_3})}{\omega_{\ell m}}\,,
\end{split}
\end{equation}
where the coefficients are listed in Table \ref{TabFitting}, and $j$ is the spin of the remnant.
For mergers of nonspinning BHs, $j$ is only a function of the mass ratio $q=M_1/M_2$ ($q\ge1$) which can be approximated as  $j(q)=\eta\left(2\sqrt{3}-3.5171\eta+2.5763\eta^2\right)$ \cite{Barausse:2009uz}, and the fitting formulas of $A_{\ell m}$ are given in Refs. \cite{Baibhav:2018rfk,Baibhav:2017jhs}.
Here $\eta=q/(1+q)^2$ is the symmetric mass ratio.

\begin{table}
\begin{tabular}{ccccccc}
\hline\hline
($\ell,m$) & $f_1$ & $f_2$ & $f_3$ & $q_1$ & $q_2$ & $q_3$    \\
\hline
(2,2) & 1.5251 & -1.1568 & 0.1292 &  0.7000 & 1.4187 & -0.4990\\
(3,3) & 1.8956 & -1.3043 & 0.1818 &  0.9000 & 2.3430 & -0.4810\\
(2,1) & 0.6000 & -0.2339 & 0.4175 & -0.3000 & 2.3561 & -0.2277\\
(4,4) & 2.3000 & -1.5056 & 0.2244 &  1.1929 & 3.1191 & -0.4825\\
\hline
\end{tabular}
\caption{The coefficients \cite{Berti:2005ys} in Eq. \eqref{EqOmegaTau}.}
\label{TabFitting}
\end{table}

\begin{figure}
\includegraphics[width=0.9\columnwidth]{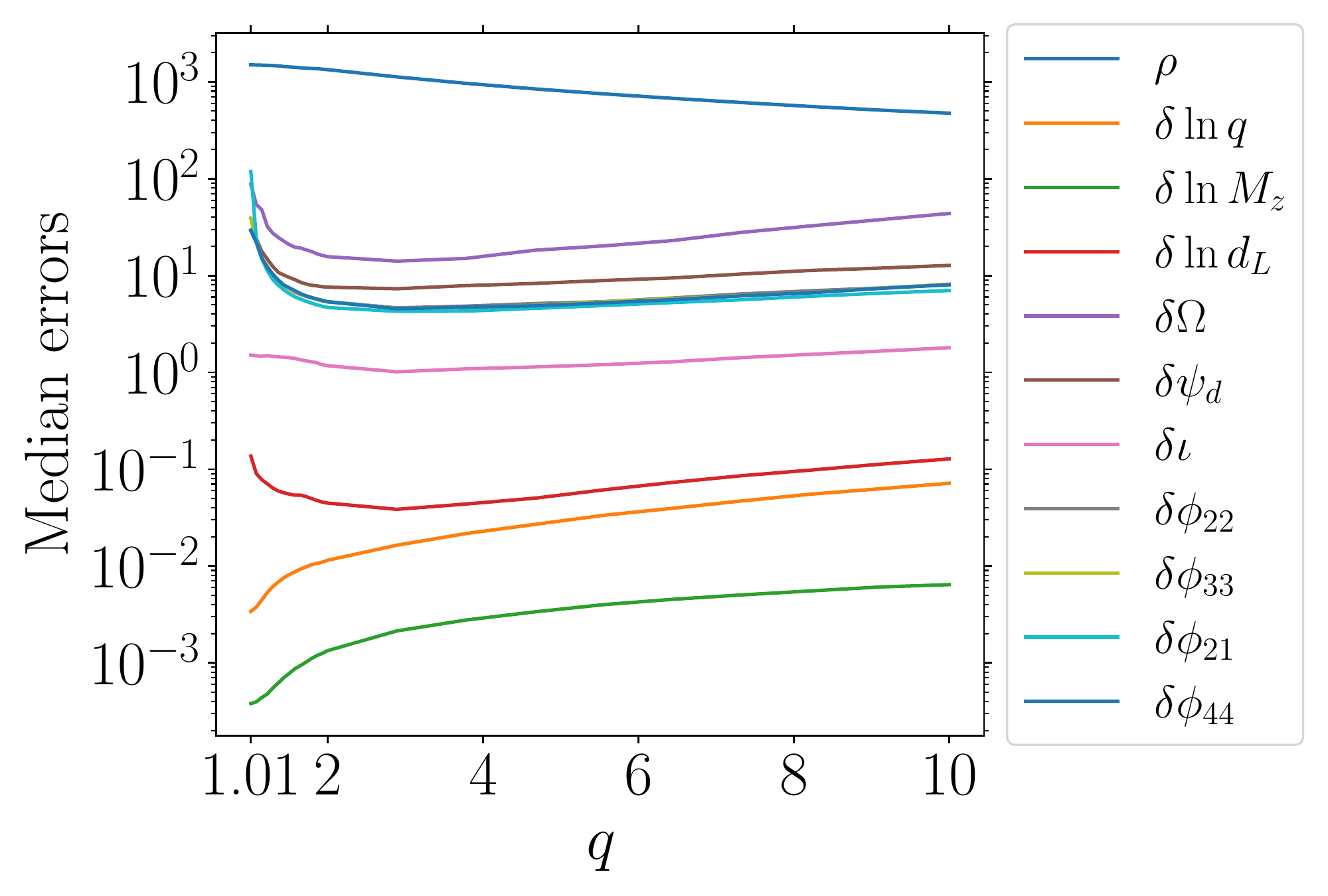}
\caption{The effect of the mass ratio on the median errors of the parameters and the source localization with TianQin for the binary with the total mass $M=10^6\ M_\odot$ at the redshift $z=1$.
}
\label{Figq}	
\end{figure}

In this paper, we take $q=2$.
In fact, for different $q$ ($1< q\le 10$), the difference of the results for the parameter estimation and source localization is mostly within one order of magnitude as shown in Fig. \ref{Figq}.

\subsection{Polarization tensors}
In the heliocentric coordinate $\{\hat{i},\hat{j},\hat{k}\}$, the GW coordinate basis vectors $\{\hat{m},\hat{n},\hat{o}\}$ are determined by the source location $(\theta_s,\phi_s)$ and the polarization angle $\psi_s$ as
\begin{equation}\label{EqGWHelioVectors}
\{\hat{m},\hat{n},\hat{o}\}=\{\hat{i},\hat{j},\hat{k}\}\times R_z\left(\phi_{s}-\pi\right) R_y\left(\pi-\theta_{s}\right) R_z\left(\psi_s\right),
\end{equation}
where $\hat{o}$ is the propagating direction of GWs, and $R_x$, $R_y$, and $R_z$ are Euler rotation matrices given by Eq. \eqref{EqEuler}.

In general relativity, there are two polarizations $A=+,\times$.
With the help of polarization tensors $e^A_{ij}$,
\begin{equation}\label{EqPolar}
e^{+}_{ij}=\hat{m}_i\hat{m}_j-\hat{n}_i\hat{n}_j,\qquad  e^{\times}_{ij}=\hat{m}_i\hat{n}_j+\hat{n}_i\hat{m}_j,
\end{equation}
we can decompose GWs into two polarizations $h_{ij}=\sum_{A=+,\times} h_A e^A_{ij}$.

\subsection{The detector signal}

To use FIM to estimate parameters, we need the frequency-domain signal $s(f)$ in the detector.
An analytical expression for $s(f)$ will help to speed up the computation and improve the precision.
In Appendix \ref{AppAnalytical}, we present the analytical formulas used in this paper.

\begin{figure}
\includegraphics[width=0.6\columnwidth]{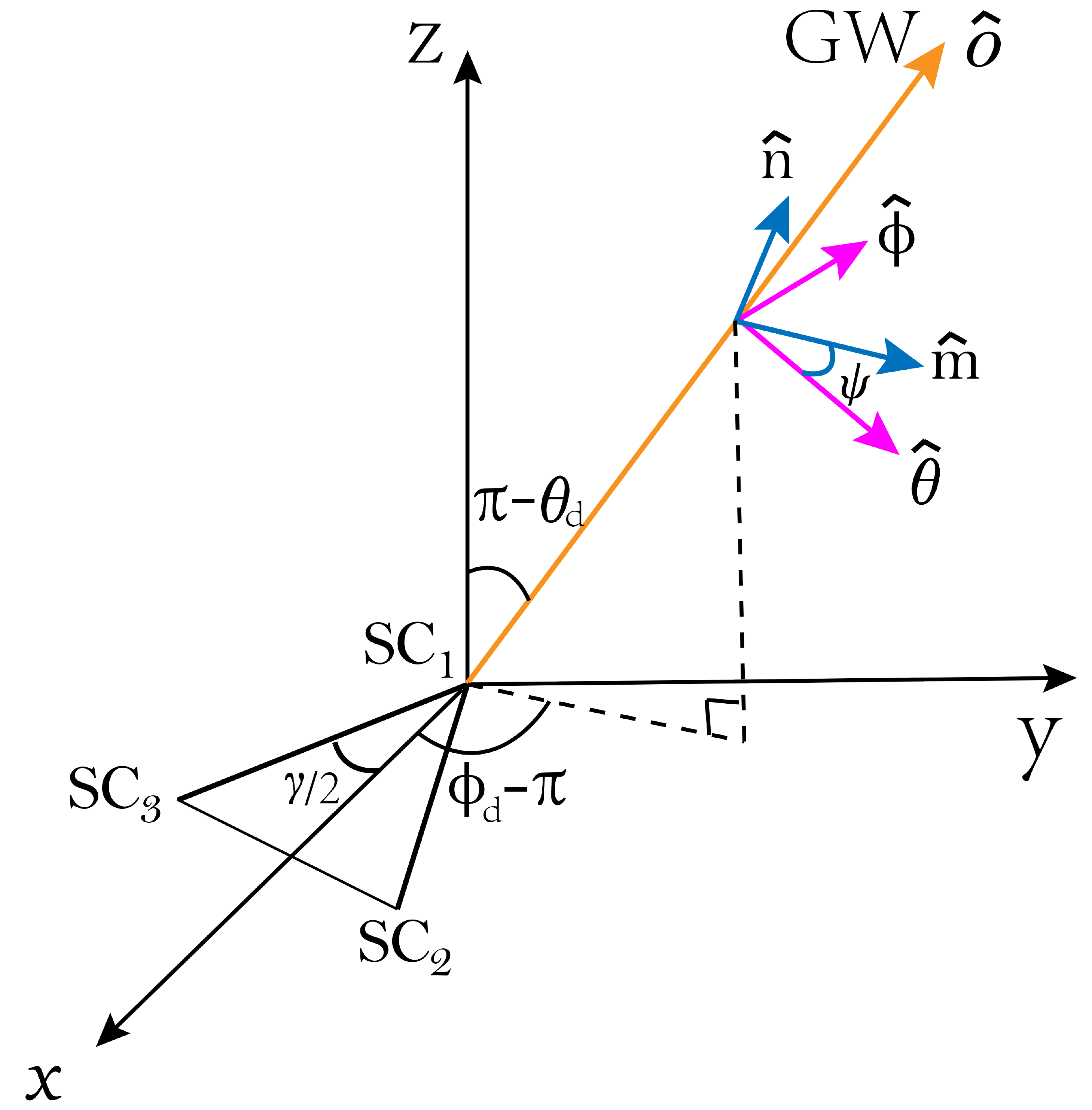}
\caption{
The detector coordinate and the configuration of the detector with the opening angle $\gamma=\pi/3$.
}
\label{FigDet}
\end{figure}

At the ringdown stage, the damping time of GWs is normally within one day. 
Since space-based GW detectors take one year to orbit around the Sun, 
we treat the Doppler shift $\exp\left[-2\pi f \hat{o}\cdot\vec{r}_0/c\right]$, where $\vec{r}_0$ is the position of the center of mass of the detector in the heliocentric coordinate, as a constant at the ringdown stage.
For convenience, we work in the detector coordinate as shown in Fig. \ref{FigDet}.
Although we put the detector in the $x$-$y$ plane, it is straightforward to obtain the direction that the detector plane points to in the heliocentric coordinate.
In the heliocentric coordinate, if the heliocentric detector (such as LISA and Taiji) is at $(\theta_s,\phi_s)=(\pi/2,\phi_0)$, then the normal vector of its detector plane will point to $(\theta_s,\phi_s)=(\pi/3,\phi_0+\pi)$.

In the detector coordinate, the polarization tensors are given by
\begin{equation}\label{EqRingPolar}
\begin{split}
&e^{+}_{ij}=\hat{m}_i\hat{m}_j-\hat{n}_i\hat{n}_j,\qquad  e^{\times}_{ij}=\hat{m}_i\hat{n}_j+\hat{n}_i\hat{m}_j,\\
&\{\hat{m},\hat{n},\hat{o}\}=R_z\left(\phi_{d}-\pi\right) R_y\left(\pi-\theta_d\right) R_z\left(\psi_d\right),
\end{split}
\end{equation}
where $(\theta_d,\phi_d,\psi_d)$ are source parameters in the detector coordinate,
which are determined by $(\theta_s,\phi_s,\psi_s)$ through Eqs. \eqref{EqEuler}-\eqref{DetParams}.

The configurations of space-based GW detectors are generally equilateral triangles.
We can model every detector of this kind as a combination of two independent LIGO-like detectors (``I" and ``II") with the opening angle $\gamma=\pi/3$.
Thus the signal in the detector II for a source at ($\theta_d,\phi_d$) is equivalent to the signal in the detector I for the same source but at ($\theta_d,\phi_d-2\pi/3$).

\subsubsection{Geocentric detectors}
Geocentric detectors orbit the Earth and further rotate around the Sun together with the Earth.
We take TianQin as an example, whose detector
plane faces to the source RX J0806.3+1527 at ($\theta_{tq} =
94.7^\circ$, $\phi_{tq} = 120.5^\circ$) \cite{Israel:2002gq,Barros:2004er,Roelofs:2010uv,Esposito:2013vja,Kupfer:2018jee}.

In the case of $10^5\ M_\odot \leq M_z \leq 10^7 \ M_\odot$, the damping time of the ringdown signals is within 10 minutes.
Thus we ignore the rotation of TianQin in this case.
The frequency-domain detector signal is
\begin{equation}\label{EqTQsf1}
s(f)=\sum_{A=+,\times} \left[D_u^A \mathcal{T}(f,\hat{u}\cdot\hat{o})-D_v^A \mathcal{T}(f,\hat{v}\cdot\hat{o})\right]h_{A}(f),
\end{equation}
where 
\begin{equation}\label{EqDefDuv}
D_u^A=\frac{1}{2}\hat{u}^i \hat{u}^j e^A_{ij},\; D_v^A=\frac{1}{2}\hat{v}^i \hat{v}^j e^A_{ij},
\end{equation}
the unit vectors of the detector's two arms are 
\begin{equation}\label{Equv1}
\begin{split}
\hat{u}=&\left[\cos\left(\frac{\gamma}{2}\right), -\sin\left(\frac{\gamma}{2}\right),0\right],
\\
\hat{v}=&\left[\cos\left(\frac{\gamma}{2}\right),\sin\left(\frac{\gamma}{2}\right),0\right],
\end{split}
\end{equation}
and $e^A_{ij}$ is the polarization tensor, given by Eq. \eqref{EqRingPolar}.
Combining Eqs. \eqref{EqRingPolar} and \eqref{Equv1}, we get
\begin{equation}\label{EqDuv}
\begin{split}
D^+_u=&\frac{1}{4}\left[\left(1+\cos^2\theta_d\right)\cos(2\phi_d+\gamma)-\sin^2\theta_d\right]\cos(2\psi_d)\\
&+\frac12\cos\theta_d\sin(2\phi_d+\gamma)\sin(2\psi_d), \\
D^+_v=&\frac{1}{4}\left[\left(1+\cos^2\theta_d\right)\cos(2\phi_d-\gamma)-\sin^2\theta_d\right]\cos(2\psi_d)\\
&+\frac12\cos\theta_d\sin(2\phi_d-\gamma)\sin(2\psi_d),\\
D^\times_u=&\frac{1}{4}\left[\sin^2\theta_d-\left(1+\cos^2\theta_d\right)\cos(2\phi_d+\gamma)\right]\sin(2\psi_d)\\
&+\frac12\cos\theta_d\sin(2\phi_d+\gamma)\cos(2\psi_d),\\
D^\times_v=&\frac{1}{4}\left[\sin^2\theta_d-\left(1+\cos^2\theta_d\right)\cos(2\phi_d-\gamma)\right]\sin(2\psi_d)\\
&+\frac12\cos\theta_d\sin(2\phi_d-\gamma)\cos(2\psi_d).
\end{split}
\end{equation}
The transfer function $\mathcal{T}$ is 
\begin{equation}
\label{EqTransfer}
\begin{split}
&\mathcal{T}(f,\hat{u}\cdot\hat{o})\\
&=\frac{1}{2}\left\{{\rm sinc}\left[\frac{f(1-\hat{u}\cdot\hat{o})}{2f^*}\right]\exp\left[\frac{f(3+\hat{u}\cdot\hat{o})}{2if^*}\right]\right.\\ 
&\quad \left.+\text{sinc}\left[\frac{f(1+\hat{u}\cdot\hat{o})}{2f^*}\right]\exp\left[\frac{f(1+\hat{u}\cdot\hat{o})}{2if^*}\right]\right\},
\end{split}
\end{equation}
where $\text{sinc}(x)=\sin x/x$, $f^*=c/(2\pi L)$ is the transfer frequency of the detector, $c$ is the speed of light, and $L$ is the arm length of the detector.
The frequency-domain GW signal $h_A(f)$ is
\begin{equation}\label{Eqhf}
\begin{split}
h_+(f)&=\frac{M_z}{d_L}\sum_{\ell,m}A_{\ell m}Y^{\ell m}_+(\iota) I_a\left(\omega_{\ell m},\tau_{\ell m},\phi_{\ell m}\right),\\
h_\times(f)&=\frac{M_z}{d_L}\sum_{\ell,m}A_{\ell m}Y^{\ell m}_\times(\iota) I_a\left(\omega_{\ell m},\tau_{\ell m},\phi_{\ell m}-\frac{\pi}{2}\right),
\end{split}
\end{equation}
where $I_a$ is given by Eq.  \eqref{EqIa}.
For the detector II, the analytical frequency-domain detector signal is given by the replacement $\phi_d \to \phi_d-2\pi/3$ in Eq. \eqref{EqTQsf1}.

In the case of $10^7\ M_\odot \leq M_z \leq 1.47\times10^9 \ M_\odot$, the damping time of the ringdown signals is from 10 minutes to one day with the frequencies of the four strongest modes within a few mHz.
We choose to treat the Doppler shift $\exp\left[-2\pi if \hat{o}\cdot\vec{r}_1/c\right]$, which comes from the time shift between $\vec{r}_1$ (the position of $SC_1$) and the coordinate origin, as a constant, 
because $f$ is within mHz, $|\vec{r}_1|/c=1/3$ s, 
and the variation of $\vec{r}_1$ is small.
We approximate $\mathcal{T}(f)\approx 1$ because the GW frequency is much less than the transfer frequency of TianQin, and take into account the rotation of TianQin,
so $\hat{u}$ and $\hat{v}$ are 
\begin{equation}\label{Equv2}
\begin{split}
\hat{u}=&\left[\cos\left(\omega_{tq}t-\frac{\gamma}{2}\right), \sin\left(\omega_{tq}t-\frac{\gamma}{2}\right),0\right],\\
\hat{v}=&\left[\cos\left(\omega_{tq}t+\frac{\gamma}{2}\right),\sin\left(\omega_{tq}t+\frac{\gamma}{2}\right),0\right],
\end{split}
\end{equation}
where $\omega_{tq}=2\pi/T_{tq}=1.99\times10^{-5}$ Hz is the rotation frequency of TianQin.
For the detector I, the analytical frequency-domain detector signal is
\begin{widetext}
\begin{equation}\label{EqTQsf2}
\begin{split}
s(f)=&\Big[-\frac{\sin\gamma}{2}(1+\cos^2\theta_d)\cos(2\psi_d)I_b\left(\phi_d, \omega_{\ell m}, \tau_{\ell m}, \phi_{\ell m}\right)\\
&+\sin\gamma\cos\theta_d\sin(2\psi_d)I_b\left(\phi_d+\frac{\pi}{4}, \omega_{\ell m}, \tau_{\ell m}, \phi_{\ell m}\right)\Big]
\frac{M_z A_{\ell m}}{d_L}Y^{\ell m}_+(\iota)\\
&+\Big[\frac{\sin\gamma}{2}(1+\cos^2\theta_d)\sin(2\psi_d) I_b\left(\phi_d, \omega_{\ell m}, \tau_{\ell m}, \phi_{\ell m}-\frac{\pi}{2}\right) \\
&+\sin\gamma\cos\theta_d\cos(2\psi_d)I_b\left(\phi_d+\frac{\pi}{4}, \omega_{\ell m}, \tau_{\ell m}, \phi_{\ell m}-\frac{\pi}{2}\right) \Big]\frac{M_z A_{\ell m}}{d_L}Y^{\ell m}_\times(\iota).
\end{split}
\end{equation}
\end{widetext}
Here $I_b$ is given by Eq. \eqref{EqIb}.
For the detector II, the analytical frequency-domain detector signal is given by 
the replacement $\phi_d \to \phi_d-2\pi/3$ in Eq. \eqref{EqTQsf2}.

If a geocentric detector has a longer arm length than TianQin, it will have a longer rotation period and a lower transfer frequency than TianQin.
Thus we only need to increase the boundary redshifted mass (we choose $10^7\ M_\odot$ for TianQin) in the above two cases.

\subsubsection{Heliocentric detectors}

The heliocentric detector rotates around the Sun in the orbit of the Earth, with a fixed period of one year.
We take LISA as an example, in the case of $10^5\ M_\odot \leq M_z \leq 1.47\times10^9 \ M_\odot$, since the rotation frequency of LISA $\omega_{lisa}=1.99\times10^{-7}$ Hz is extremely small, we ignore the rotation of LISA.
The arm vectors $\hat{u}$ and $\hat{v}$ are the same as Eq. \eqref{Equv1}, 
and the expressions for the frequency-domain detector signals are  the  same as Eqs. \eqref{EqTQsf1}-\eqref{Eqhf}.
This does not mean that LISA and TianQin will have the same detector signal, 
because the two detectors have different detector parameters $(\theta_d,\phi_d,\psi_d)$ for the same source $(\theta_s,\phi_s,\psi_s)$ (Appendix \ref{CoorTrans}), 
different transfer frequency, different noise, etc.

\subsection{The noise curve}

In this paper, we use the noise curve \cite{Cornish:2018dyw}
\begin{equation}\label{EqPn}
P_n(f) = \frac{S_x}{L^2}  + \frac{2[1+\cos^2(f/f^*)]S_a}{(2\pi f)^4 L^2}\left[1+ \left(0.4\ {\rm mHz}/f\right)^2 \right],
\end{equation}
where $S_x$ is the position noise, $S_a$ is the acceleration noise, $L$ is the arm length,
$f^*=c/(2\pi L)$ is the transfer frequency of the detector.
For LISA, $S_x = (1.5 \times 10^{-11} \ {\rm m})^2  \ {\rm Hz}^{-1}$, $S_a = (3 \times 10^{-15} \ {\rm m}\, {\rm s}^{-2})^2 \ {\rm Hz}^{-1}$, $L=2.5\times10^9$ m  and $f^*=19.09$ mHz \cite{Audley:2017drz}.
For TianQin, $S_x = (10^{-12}\ {\rm m})^2  \ {\rm Hz}^{-1}$, $S_a = (10^{-15} \ {\rm m}\, {\rm s}^{-2})^2\  {\rm Hz}^{-1}$, $L=\sqrt{3}\times10^8$ m and $f^*=0.2755$ Hz \cite{Luo:2015ght}.
For Taiji, $S_x = (8\times10^{-12} \ {\rm m})^2  \ {\rm Hz}^{-1}$, $S_a = (3\times10^{-15} \ {\rm m}\, {\rm s}^{-2})^2\ {\rm Hz}^{-1}$, $L=3\times10^9$ m and $f^*=15.90$ mHz \cite{Ruan:2020smc}.

For LISA and Taiji, we also add the confusion noise \cite{Cornish:2018dyw}
\begin{equation}\label{EqSc}
\begin{split}
S_c(f) = &\frac{2.7\times10^{-45} f^{- 7/3}}{1+0.6(f/0.01909 )^{2}}\, e^{-f^{0.138} - 221  f  \sin(521 f) } \\ &\times\left[1+{\rm tanh}(1680(0.00113-f))\right]  \ {\rm Hz}^{-1},
\end{split}
\end{equation}
to the noise curve.

\begin{figure}
\includegraphics[width=0.9\columnwidth]{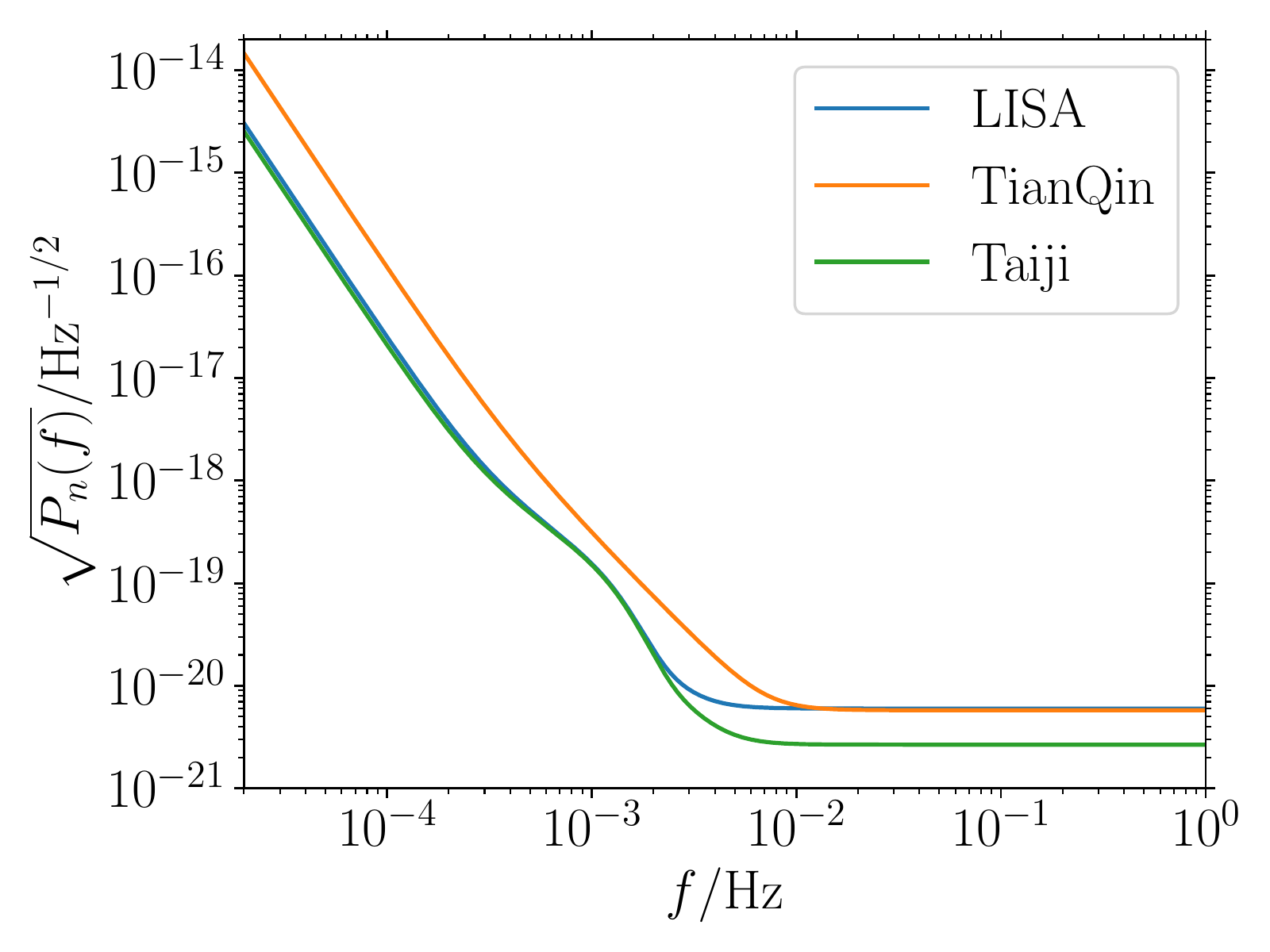}
\caption{The noise power spectra of LISA, TianQin, and Taiji.}
\label{Fignoise}	
\end{figure}

Figure \ref{Fignoise} shows the noise power spectra of  LISA, TianQin, and Taiji.

\subsection{Fisher information matrix}	

For convenience, we define the inner product of two frequency-domain signals $s_1(f)$ and $s_2(f)$ as
\begin{equation}
(s_1|s_2)=2\int_{f_{\rm in}}^{f_{\rm out}}\frac{s_1(f)s_2^*(f)+s_1^*(f)s_2(f)}{P_{n}(f)}df.
\end{equation}
The SNR $\rho$ is simply defined as
\begin{equation}
\rho^2=(s|s).
\end{equation}
For a detected source with a significant SNR  (a threshold of $ \rho\ge 8$), we can use the FIM method to estimate its parameters, which is defined as
\begin{equation}
\label{EqGammaij}
\begin{split}
\Gamma_{ij}=&\left(\frac{\partial s(f)}{\partial {\bm \xi}_i}\left|\frac{\partial s^*(f)}{\partial {\bm \xi}_j}\right.\right),
\end{split}
\end{equation}
where ${\bm \xi}=\left\{q,M_z,d_L,\theta_d,\phi_d,\psi_d,\iota,\phi_{22},\phi_{33},\phi_{21},\phi_{44}\right\}$ spans the  11-dimensional parameter space.
Although the ringdown phases $\phi_{\ell m}$ are related to the source parameters and the specific process, the relationship is not well known.
Thus we treat $\phi_{\ell m}$ as four independent parameters.

The covariance matrix of these parameters is
\begin{equation}
\sigma_{ij}=\left\langle\Delta\xi^i\Delta\xi^j\right\rangle\approx (\Gamma^{-1})_{ij}.
\end{equation}
The angular uncertainty of the sky localization is evaluated as
\begin{equation}
\Delta \Omega_s\equiv2\pi\sin\theta_d
\sqrt{\sigma_{\theta_d\theta_d}\sigma_{\phi_d\phi_d}-\sigma^2_{\theta_d\phi_d}}\,,
\end{equation}
so the probability that the source lies outside an error ellipse enclosing the solid angle $\Delta\Omega$ is simply $e^{-\Delta\Omega/\Delta\Omega_s}$.

\section{parameter estimation and source localization}	
\label{SecResults}

It is hard to control the noise of space-based detectors below the frequency $\sim2\times10^{-5}$ Hz \cite{Baibhav:2020tma}, 
so we take $2\times10^{-5}$ Hz as the lower cutoff frequency.
For BH binaries with the redshifted total mass $M_z\ge10^9\ M_\odot$, 
$f_{22}$ and $f_{21}$ are out of the frequency band of space-based detectors, 
thus we do not consider binaries with the total mass $M\ge10^9\ M_\odot$.
For ringdown signals, we set $f_{\rm in}=\max\left(0.5f_{21}, 2\times10^{-5}\ {\rm Hz}\right)$ and $f_{\rm out}=2f_{44}$.
Since higher frequencies correspond to higher overtones and higher harmonics, 
which are not used in our computation,
we choose this upper limit $f_{\rm out}$ in the integration.
The lower limit in the integration stands for the starting frequency of the ringdown stage, 
and we set it to be $0.5f_{22}$ in our computation.

\begin{figure}
\includegraphics[width=0.9\columnwidth]{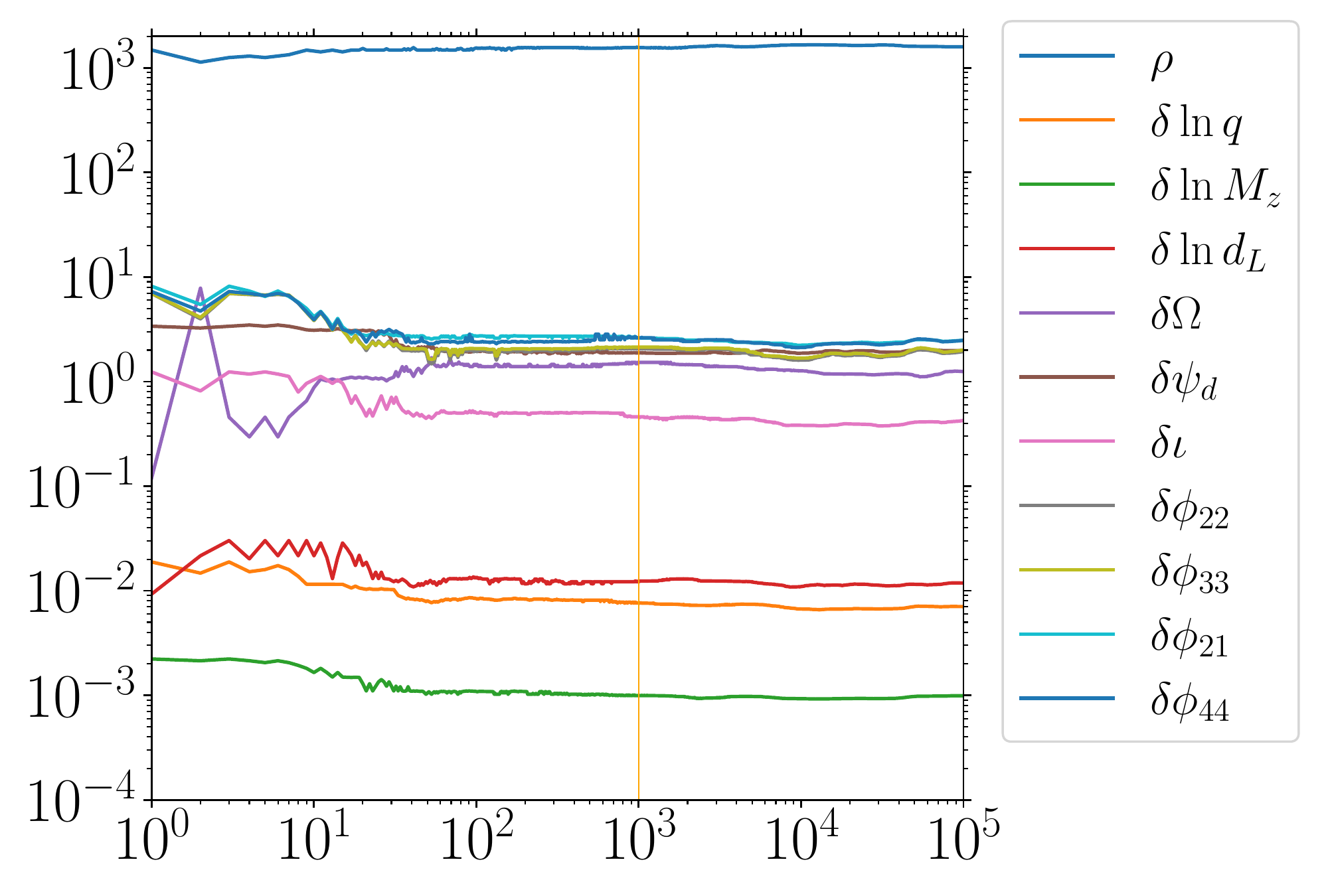}
\caption{The effect of the number of simulated sources on the median errors of the parameter estimation and source localization with LISA for the binary with $10^6\ M_\odot$ at $z=1$.
The vertical orange solid line represents the number of simulated sources that we choose to obtain the median error of each parameter.}
\label{FigNum}	
\end{figure}

In this section, for each binary with the same total mass and redshift, 
we use Monte Carlo simulation to generate 1000 sources and obtain the median error of each parameter.
We also check the effect of the number of simulated sources on the median errors and the results are shown in Fig. \ref{FigNum}. 
We see that the results are almost the same if the number of simulated sources is larger than 100, so we choose to simulate 1000 sources.
From Eqs. \eqref{EqYlm}, \eqref{EqTQsf1}, \eqref{EqDuv}, and \eqref{EqTQsf2}, 
we see that there exists a transformation of extrinsic parameters yielding an exact degeneracy,
called reflected sky position (for a reflection with respect to the detector plane) \cite{Marsat:2020rtl},
\begin{equation}
\label{EqDegeneracy1}
\begin{split}
    \theta_d\to\pi-\theta_d,\\
    \iota\to\pi-\iota,\\
    \psi_d\to\pi-\psi_d.
\end{split}
\end{equation}
Thus, in general, there are two degenerate positions in the sky in the parameter estimation with ringdown signals.  
Moreover, in the low-frequency limit, due to ${\cal T}\to 1$, the constraints on $\hat{u}\cdot\hat{o}$ and $\hat{v}\cdot\hat{o}$ in the transfer function become weak, leading to another transformation \cite{Marsat:2020rtl},
\begin{equation}\label{EqDegeneracy2}
\begin{split}
    &\phi_d\to\phi_d+\frac{k\pi}{2}\quad {\rm mod\ 2\pi},\\
    &\psi_d\to\psi_d+\frac{k\pi}{2}\quad {\rm mod\ \pi},
\end{split}
\end{equation}
where $k=0,1,2,3$.
The new transformation implies that, in the parameter estimation with ringdown signals from supermassive BH binaries, there are eight degenerate positions in the sky in the low-frequency limit. 
Note that the two transformations \eqref{EqDegeneracy1} and \eqref{EqDegeneracy2} 
lead to a multimodal distribution in the Bayesian analysis discussed in the next section,
but they are missed in the FIM analysis.
The parameter distribution of simulated sources is shown in Appendix \ref{AppSource}.

\begin{figure}
\includegraphics[width=0.6\columnwidth]{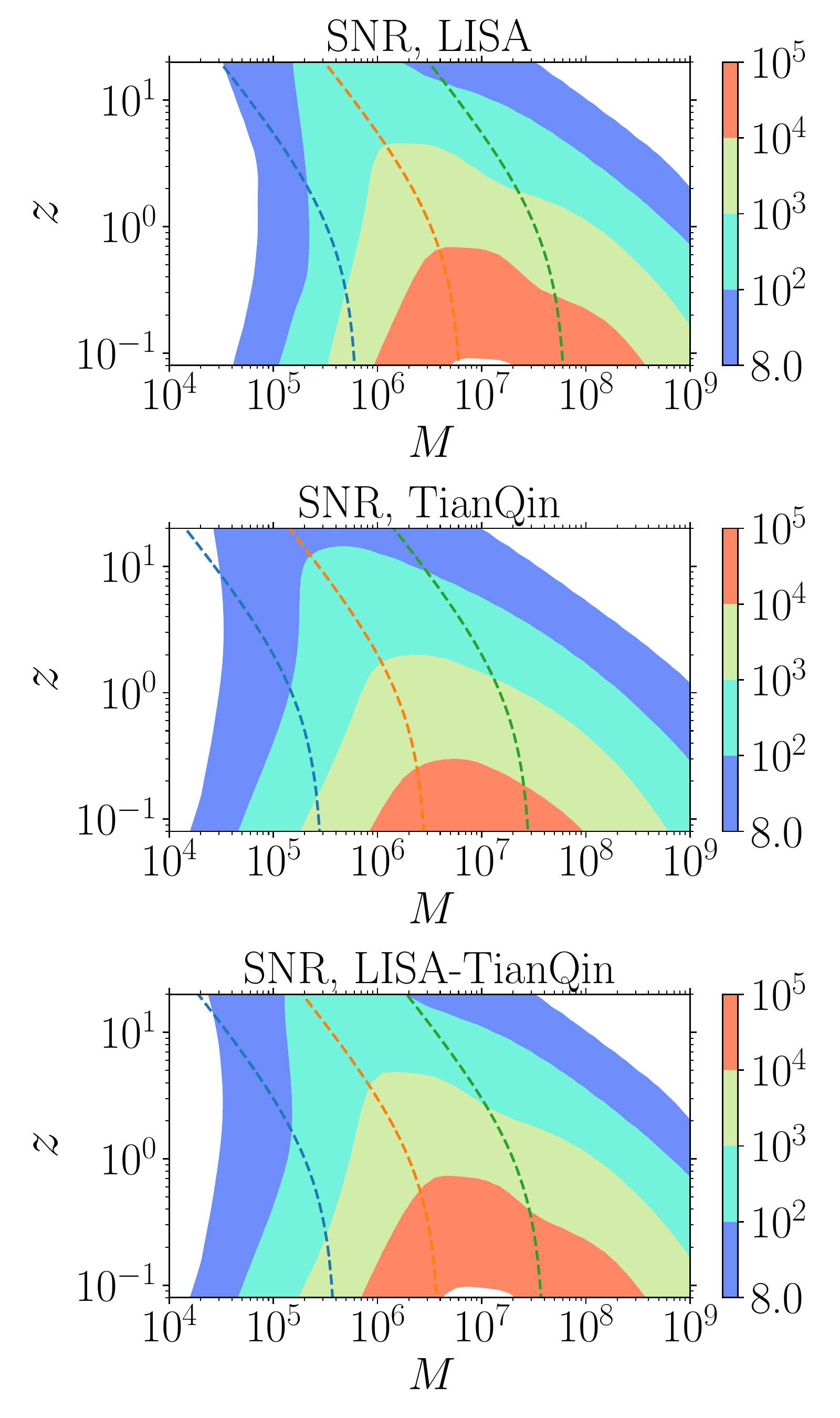}
\caption{The median SNRs of TianQin and LISA with ringdown signals from binaries with different total masses and different redshifts. 
For LISA, the blue, orange, and green dashed lines correspond to the redshifted masses $M_z=6.5\times10^5\ M_\odot$, $6.5\times10^6\ M_\odot$ and $6.5\times10^7\ M_\odot$, respectively.
For TianQin, the blue, orange, and green dashed lines correspond to the redshifted masses $M_z=3\times10^5\ M_\odot$, $3\times10^6\ M_\odot$ and $3\times10^7\ M_\odot$, respectively.
For LISA-TianQin network, the blue, orange, and green dashed lines correspond to the redshifted masses $M_z=4\times10^5\ M_\odot$, $4\times10^6\ M_\odot$ and $4\times10^7\ M_\odot$, respectively.
}
\label{FigSNR}	
\end{figure}

\begin{figure}
\includegraphics[width=\columnwidth]{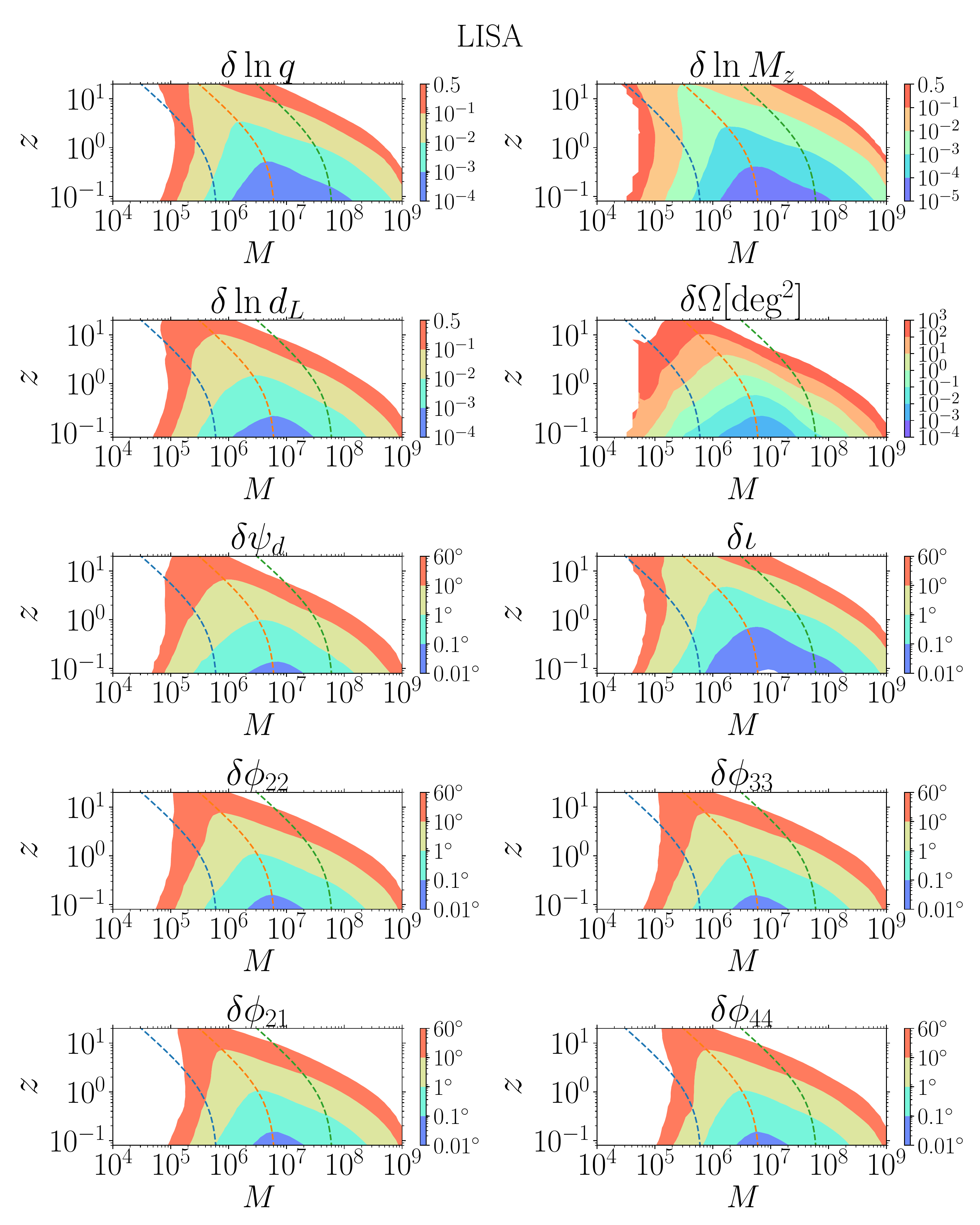}
\caption{The median errors of the parameter estimation and source localization of LISA with ringdown signals from binaries with different masses and different redshifts.
The blue, orange, and green dashed lines correspond to the redshifted masses $M_z=6.5\times10^5\ M_\odot$, $6.5\times10^6\ M_\odot$ and $6.5\times10^7\ M_\odot$, respectively.}
\label{FigLISA}	
\end{figure}

\begin{figure}
\includegraphics[width=\columnwidth]{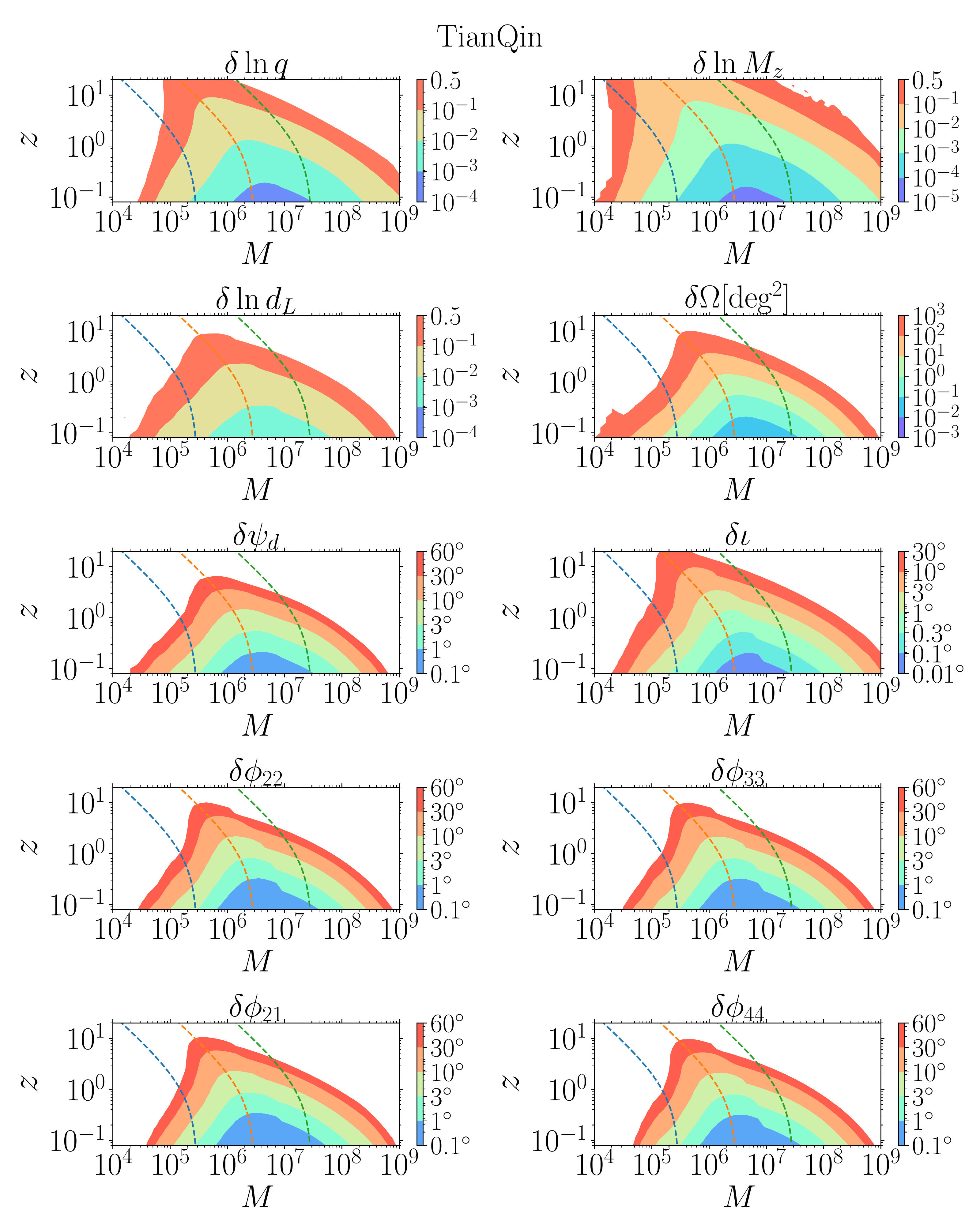}
\caption{The median errors of the parameter estimation and source localization of TianQin with ringdown signals from binaries with different masses and different redshifts. 
The blue, orange, and green dashed lines correspond to the redshifted masses $M_z=3\times10^5\ M_\odot$, $3\times10^6\ M_\odot$ and $3\times10^7\ M_\odot$, respectively.
In the case of $M_z\ge10^7\ M_\odot$, we take the rotation of TianQin into account and adopt the low-frequency approximation.}
\label{FigTQ}	
\end{figure}

\begin{figure}
\includegraphics[width=\columnwidth]{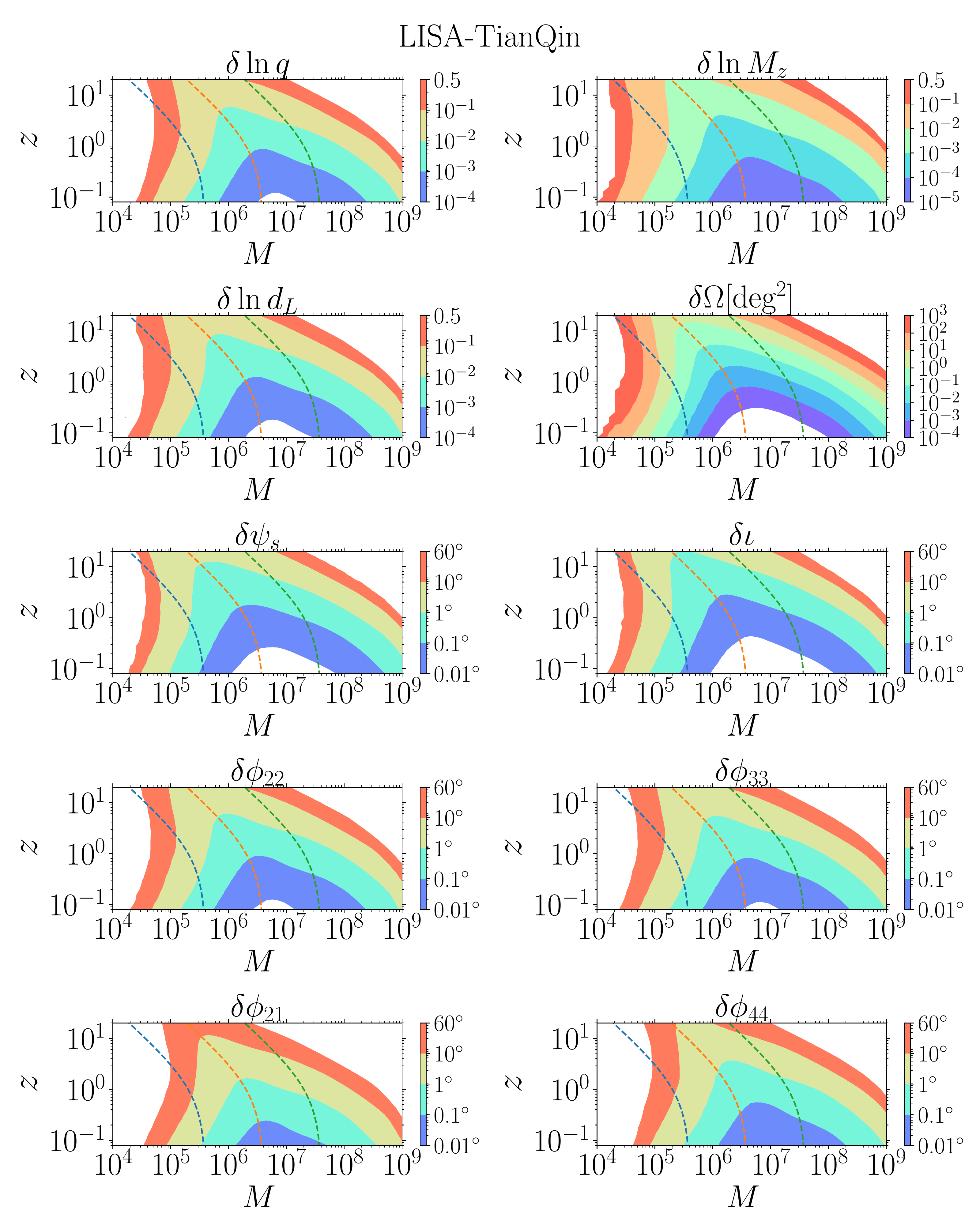}
\caption{The median errors of the parameter estimation and source localization of the network of LISA and TianQin with ringdown signals from binaries with different masses and different redshifts.
The blue, orange, and green dashed lines correspond to the redshifted masses $4\times10^5\ M_\odot$, $4\times10^6\ M_\odot$ and $4\times10^7\ M_\odot$, respectively.
In the case of $M_z\ge10^7\ M_\odot$, we take the rotation of TianQin into account and adopt the low-frequency approximation.}
\label{FigLISATQ}
\end{figure}

\begin{figure*}
\includegraphics[width=0.9\textwidth]{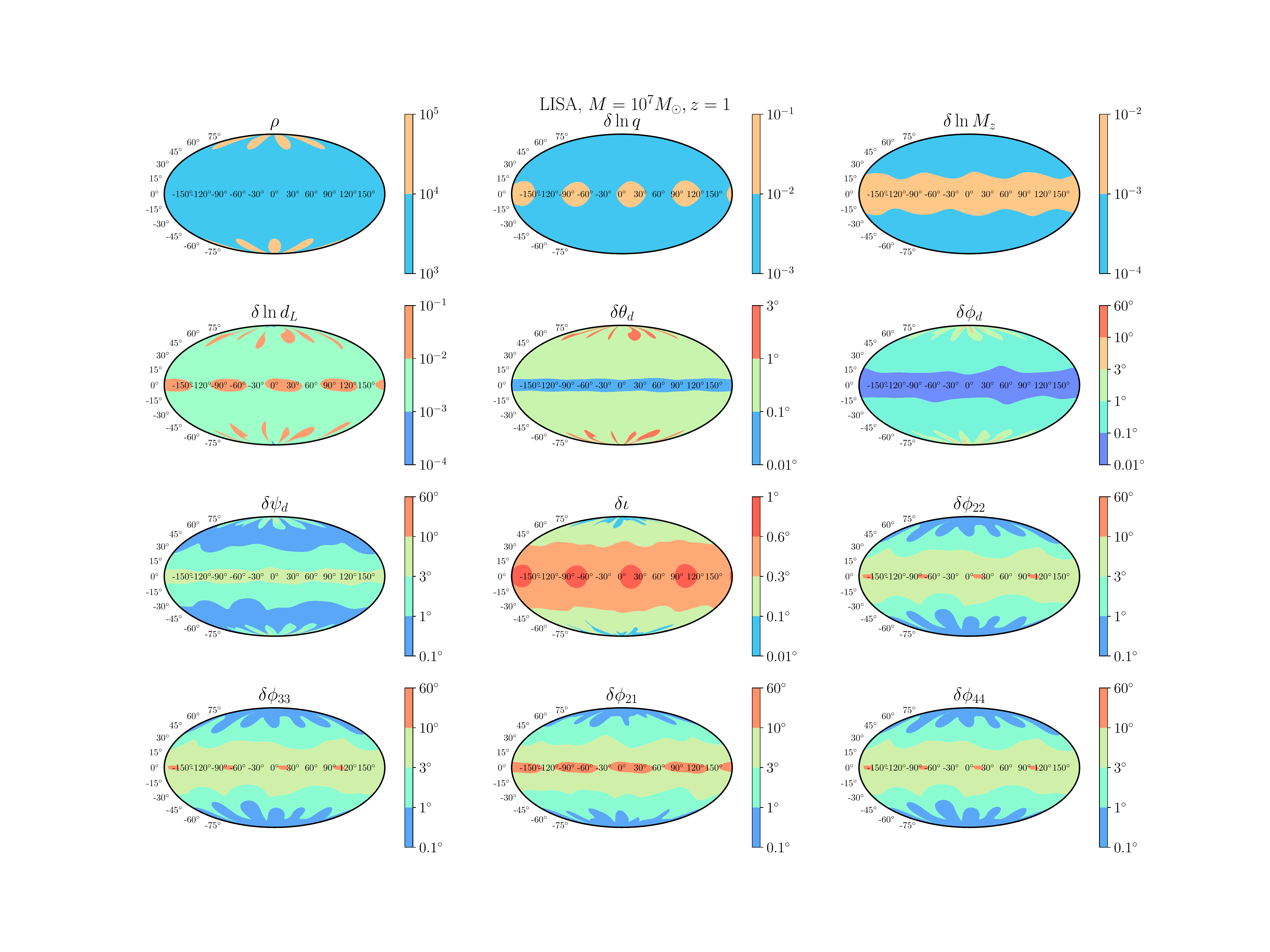}
\caption{The sky map of the parameter estimation and source localization with LISA for the binary with the total mass $M=10^7\ M_\odot$ at $z=1$.
Note that LISA is put in the equatorial plane.
The other source parameters are $q=2,\psi_d=60^\circ,\iota=45^\circ$, and $\phi_{\ell m}=0^\circ$.}
\label{FigSkyLISA}	
\end{figure*}

\begin{figure*}
\includegraphics[width=0.9\textwidth]{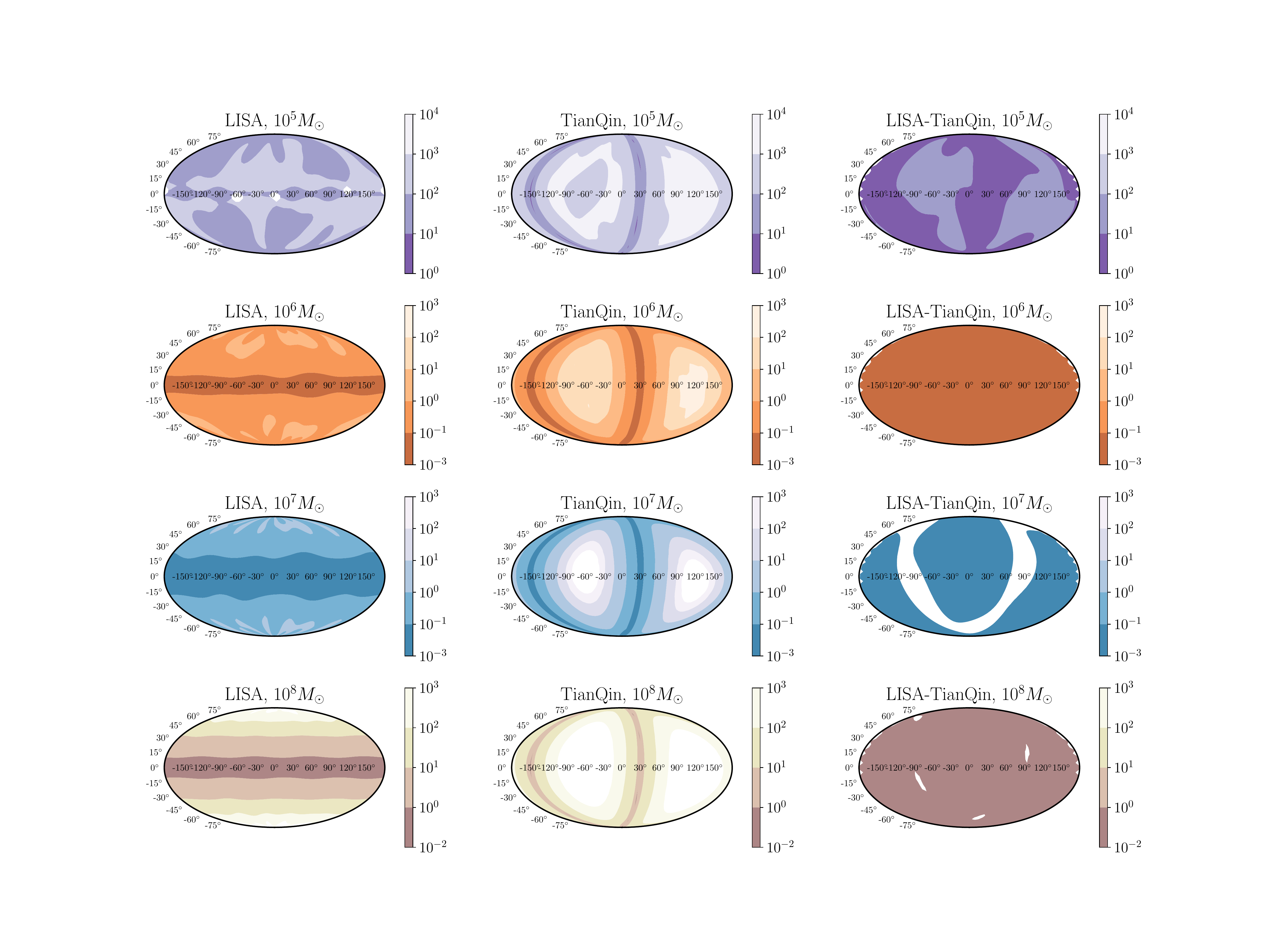}
\caption{The sky map of source localization with LISA, TianQin, and their combined network for binaries with total masses $10^5 \ M_\odot$, $10^6 \ M_\odot$, $10^7 \ M_\odot$ and $10^8 \ M_\odot$ at $z=1$.
The other source parameters are $q=2,\psi_d=\pi/3,\iota=\pi/4$, and $\phi_{\ell m}=0$.}
\label{FigNetwork}	
\end{figure*}

Using a typical source with $M=10^7\ M_\odot$ and $z=1$, 
we analyze the dependence of these errors on the sky position.
We also explore the capability of source localization for different detectors and their combined network.
The median SNR for the simulated sources is shown in Fig. \ref{FigSNR} and the results for the parameter estimation are shown in Figs. \ref{FigLISA}, \ref{FigTQ}, \ref{FigLISATQ}, \ref{FigSkyLISA}, \ref{FigNetwork}, and Tables \ref{Tabz1} and \ref{Tabz3}.
The results with Taiji are similar to those with LISA.

Figure \ref{FigLISA} shows the median errors of the parameter estimation and source localization of LISA with ringdown signals from binaries with different total masses and different redshifts,
and Fig. \ref{FigTQ} shows the median errors of the parameter estimation and source localization of TianQin.
For BH binaries with the total mass $M\ge 10^4\ M_\odot$, as the total mass increases, the SNR of its ringdown signal will exceed the threshold $\rho=8$,
the estimation errors of $\delta\ln q$, $\delta\ln M_z$ and $\delta\ln d_L$ will be within 0.5, 
the estimation errors of the angles and phases will be within $60^\circ$, 
and the source localization will be within $1000$ $\text{deg}^2$.
If its total mass $M\ge10^5\ M_\odot$, 
in most cases, the SNR of its ringdown signal will be larger than 100,
the estimation errors of $\delta\ln q$ and $\delta\ln d_L$ will be within 0.1, 
the estimation error of $\delta\ln M_z$ will be within 0.01, 
the estimation errors of the angles and phases will be within $10^\circ$, 
and the source localization will be within $10$ $\text{deg}^2$.
If its total mass $M\ge 10^6\ M_\odot$, 
the SNR can exceed $\sim 10^3$, 
the estimation errors of $\delta\ln q$ and $\delta\ln d_L$ will be within 0.01,
the estimation error of $\delta\ln M_z$ will be within 0.001,
the estimation errors of the angles and phases will be within $1^\circ$, 
and the source localization will be within $1$ $\text{deg}^2$.
Figure \ref{FigLISATQ} shows the median errors of the parameter estimation and source localization with the network of LISA and TianQin, which implies that the network can improve the parameter estimation by about one order of magnitude, and can improve the source localization by two or even three orders of magnitude compared with individual detector.
Figures \ref{FigSNR}, \ref{FigLISA}, \ref{FigTQ},  and  \ref{FigLISATQ} tell us that for BH binaries at the same distance,  
LISA has larger SNR and better parameter estimation and source localization for the BH binary with $M_z=6.5\times10^6\ M_\odot$,
TianQin has larger SNR and better parameter estimation and source localization for the BH binary with $M_z=3\times10^6\ M_\odot$,
the network of LISA and TianQin has larger SNR and better parameter estimation and source localization for the BH binary with $M_z=4\times10^6\ M_\odot$.

There are three main factors that affect the source localization.
The first factor is the relative difference of the source location between the detector I and the detector II.
We use the angle between the direction $(\theta_d,\phi_d)$ and the direction $(\theta_d,\phi_d-2\pi/3)$ to represent the difference, which reaches the maximum at $\theta_d=\pi/2$ and reaches the minimum at $\theta_d=\{0,\pi\}$.
The second factor is the transfer function $\cal{T}$.
For the ringdown frequency $0.5f^*\le f \le 5f^*$, $\cal{T}$ slightly weakens the response and dramatically improves the source localization.
For the ringdown frequency $f\ge 10f^*$, ${\cal T}$ significantly weakens the response and the estimation of all parameters.
For the ringdown frequency $f\le 0.1f^*$, ${\cal T}$ contributes little to the source localization because ${\cal T}\to1$.
Thus the transfer function of LISA can improve their source localization for binaries with the redshifted total masses $1.7\times10^5\ M_\odot\le M_z \le 1.7\times10^6\ M_\odot$.
The transfer function of Taiji can improve its source localization for the binary with the redshifted total mass $2\times10^5\ M_\odot\le M_z \le 2\times10^6\ M_\odot$.
The transfer function of TianQin can improve its source localization for the binary with the redshifted total mass $1.2\times10^4\ M_\odot\le M_z \le 1.2\times10^5\ M_\odot$.
The third factor is the different responses of the detector to different QNMs.
In one LIGO-like detector, because of $Y^{\ell \ell}_{+,\times}\propto (\sin\iota)^{\ell-2}Y^{22}_{+,\times}$, the response difference between the (2,1) mode and the (2,2) mode is bigger than those from (3,3) and (4,4) modes.
However, the difference is still not big enough for one LIGO-like detector to localize the source.
Thus, for the source localization of each space-based GW detector, we need to consider both the detector I and II.

Figure \ref{FigSkyLISA} shows the dependence of these errors on the sky position for LISA.
Note that we place LISA in the $x$-$y$ plane, i.e., the equatorial plane.
For the LISA-TianQin network, we set LISA pointing to $(\theta_s,\phi_s)=(\pi/3,0)$.
For the LISA-TianQin-Taiji network (3-network), we set Taiji pointing to $(\theta_s,\phi_s)=(\pi/3,0)$, and set LISA pointing to $(\theta_s,\phi_s)=(\pi/3,-2\pi/9)$.
Fig. \ref{FigSkyLISA} implies that the angular resolution becomes the best for sources along the detector plane,
where the relative difference of source location reaches the maximum, 
but it is the worst for sources locating perpendicular to the detector plane, 
where the relative difference of source location reaches the minimum.

The tensor response of a detector reaches the minimum at its angular bisector and the vertical direction of its angular bisector in the detector plane \cite{Liang:2019pry,Zhang:2019oet,Zhang:2020khm}, which is at the longitude of $\{-90^\circ$, $0^\circ$, $90^\circ$, $180^\circ\}$ in the equatorial plane for the detector I, and at the longitude of \{$-150^\circ$, $-60^\circ$, $30^\circ$, $120^\circ$\} in the equatorial plane for the detector II. 
The tensor response of a detector reaches the maximum at the direction perpendicular to the detector plane, which is near the two poles.
Thus, except $\{\theta_d,\phi_d\}$, 
the worst accuracy of the parameter estimation generally occurs for sources along the detector plane, where the tensor response reaches the minimum, 
and the best accuracy of the parameter estimation generally occurs for sources along the two poles, 
where the tensor response reaches the maximum.
However, as $\theta_d \to 0$, 
both $\hat{u}\cdot\hat{o}$ and $\hat{v}\cdot\hat{o}$ in $\cal{T}$ go to 0, 
so $\cal{T}$ contributes little to the estimation of $\phi_d$.
Furthermore, from Eq. \eqref{EqDuv}, 
we see that $\phi_d$ and $\psi_d$ degenerates into one parameter $\phi_d-\psi_d$ when $\theta_d=0$ due to the coupling.
Thus the estimation of $\psi_d$ becomes the worst for binaries along the two poles. 

Figure \ref{FigNetwork} shows the dependence of the source localization on the sky position with the ringdown signal for LISA, TianQin, and their combined network.
The combined network of two detectors not only improves the localization accuracy but also makes the sky map more uniform.
Tables \ref{Tabz1} and \ref{Tabz3} show the median localization errors of different detectors with ringdown signals from binaries with total masses $\{10^5,10^6,10^7,10^8,10^9\}\ M_\odot$ at the redshift $z=1$ and the redshift $z=3$ respectively.
From the two tables, we see that Taiji has better source localization than LISA and TianQin due to its lower noise curve for massive BH binaries and lower transfer frequency for supermassive BH binaries.
As the total mass of the BH binary increases, 
the improvement in the source localization is one to three orders for the  LISA-TianQin network compared with individual detector.
The 3-network improves the sky localization only a few times than the LISA-TianQin network.

\begin{table}
\centering	
\resizebox{0.8\columnwidth}{!}{
\begin{tabular}{|c|c|c|c|c|c|}
\hline
$M(M_\odot)$& LISA & TianQin & Taiji & LISA-TianQin & 3-network \\
\hline
$10^5$& 139.87 & 2122.59 & 30.00 & 14.21 & 4.24  \\
\hline
$10^6$& 1.53 & 10.97    &  0.243 & 0.00645 & 0.00276 \\
\hline
$10^7$& 1.00 & 11.36    & 0.544 & 0.00303 & 0.000745 \\
\hline
$10^8$& 160.95 & 1180.32 & 113.61 & 0.108 & 0.0211 \\
\hline
$10^9$& $>10^5$ & $>10^5$  & $>10^5$ & 80.13 & 12.99 \\
\hline
\end{tabular}
}
\caption{Median localization errors of different detectors with ringdown signals from binaries with total masses $\{10^5,10^6,10^7,10^8,10^9\}\ M_\odot$ at the redshift $z=1$.}
\label{Tabz1}
\end{table}

\begin{table}
\centering	
\resizebox{0.8\columnwidth}{!}{
\begin{tabular}{|c|c|c|c|c|c|}
\hline
$M(M_\odot)$& LISA & TianQin & Taiji & LISA-TianQin & 3-network \\
\hline
$10^5$& 252.90 & 4893.65 & 45.72 & 16.26 & 5.60  \\
\hline
$10^6$& 9.09 & 62.97 & 1.78 & 0.033 &  0.0101 \\
\hline
$10^7$& 103.99 & 458.02 & 82.27 & 0.112 &  0.0415 \\
\hline
$10^8$& 18963 & $>10^5$ & 13191 & 9.50 & 1.518 \\
\hline
$10^9$& $>10^5$ & $>10^5$ & $>10^5$ & $>10^5$ & $>10^5$ \\
\hline
\end{tabular}
}
\caption{Median localization errors of different detectors with ringdown signals from binaries with total masses $\{10^5,10^6,10^7,10^8, 10^9\}\ M_\odot$ at the redshift $z=3$.}
\label{Tabz3}
\end{table}

Reference \cite{Marsat:2020rtl} analyzes two binaries with $M=4\times10^5\ M_\odot$, $q=3$ and $z=4$, using Bayesian inference method, and the localization errors are about 200 square degrees, which are consistent with the median error 80 square degrees given in Fig. \ref{FigLISA}.
Reference \cite{Katz:2020hku} analyzes five binaries with $M=2\times10^6\ M_\odot$, $q=3$
and $z=4$, using Bayesian inference method and the PhenomHM waveform with higher-order harmonic modes and aligned spins.
The localization errors of the five sources are about 0.001 square radians or 3 square degrees, which are consistent with the median error 1 square degree given in Fig. \ref{FigLISA}.
The localization errors of the binary with $M=4\times10^7\ M_\odot$, $q=5$ and $z=2$,
and the binary with $M=3\times10^5\ M_\odot$, $q=1.4$ and $z=7$,
are about 2000 and 800 square degrees respectively,
which are roughly consistent with the median error 500 square degrees given in Fig. \ref{FigLISA}.
In Ref. \cite{Baibhav:2020tma}, they use the low-frequency limit and
found that the estimation errors for $q=10$ are a few times worse than those for $q=2$ which are consistent with the results shown in Fig. \ref{Figq}.
In the low-frequency limit, the transfer function is independent of the frequency,
so there are eight degenerate sky positions in the localization contours \cite{Baibhav:2020tma}.
In this paper, we consider the frequency dependence of the transfer function and use it to improve the sky localization.
As we will see in the next section, the transfer function also helps to break the eight degeneracy \eqref{EqDegeneracy2}.

\section{Bayesian inference}
\label{SecBayesian}

To check the FIM results, we use Bayesian inference method to analyze two specific sources and compare the FIM results with those from Bayesian analysis.

Bayesian inference method is based on Bayes rule
\begin{equation}\label{EqBayes}
    p({\bm \xi}|d) = \frac{p(d|{\bm \xi})p({\bm \xi})}{p(d)},
\end{equation}
where $p({\bm \xi}|d)$ is the posterior distribution of the parameters ${\bm \xi}$, $p(d|{\bm \xi})$ is the likelihood,
\begin{equation}\label{EqLikelihood}
    p(d|{\bm \xi})=\exp\left[-\frac{1}{2}(s({\bm \xi})-d|s({\bm \xi})-d)\right],
\end{equation}
$d=s({\bm \xi_0})+n$ is the observed data for the true parameters ${\bm \xi_0}$,
$n$ is the noise generated by the noise power spectra,
$p({\bm \xi})$ is the prior distribution of the parameters ${\bm \xi}$,
and $p(d)$ is the evidence which is treated as a normalization constant.
For the two sources, we choose the sampler Dynesty \cite{Speagle:2019ivv} with ``mulit" bound and ``rwalk" sample method for nested sampling \cite{Skilling04,Skilling:2006gxv}, to obtain the posterior distribution of the parameters $\bm{\xi}$.

We choose the two sources with different SNRs.
The parameters of the first source with smaller SNR are
$q=2$, $M=10^5M_\odot$, $z=1$, $\theta_s=\phi_s=\psi_s=\phi_{\ell m}=\pi/3$, and $\iota=\pi/4$.
The number of live points of the sampler, the sampling frequency, 
and the observation time are set to be 1500,  1 Hz, and 100 s, respectively.
The parameters of the second source with larger SNR are
$q=2$, $M=10^7M_\odot$, $z=1$, $\theta_s=\phi_s=\psi_s=\phi_{\ell m}=\pi/3$, and $\iota=\pi/4$.
The number of live points of the sampler, the sampling frequency, 
and the observation time are set to be 3000, 0.1 Hz, and 10000 s, respectively.
We choose $\iota=\pi/4$ because in this case the parameter errors are close to the median errors. 
From Eq. \eqref{EqDegeneracy1}, we see that in the heliocentric coordinate, the sky position $(\theta_s,\phi_s,\psi_s,\iota)=(60^\circ,60^\circ,60^\circ,45^\circ)$ is reflected to $(67.7^\circ,72.8^\circ,7.2^\circ,135^\circ)$ for LISA, and to $(55.7^\circ,6.3^\circ,130^\circ,135^\circ)$ for TianQin.

\begin{figure}
\includegraphics[width=0.9\columnwidth]{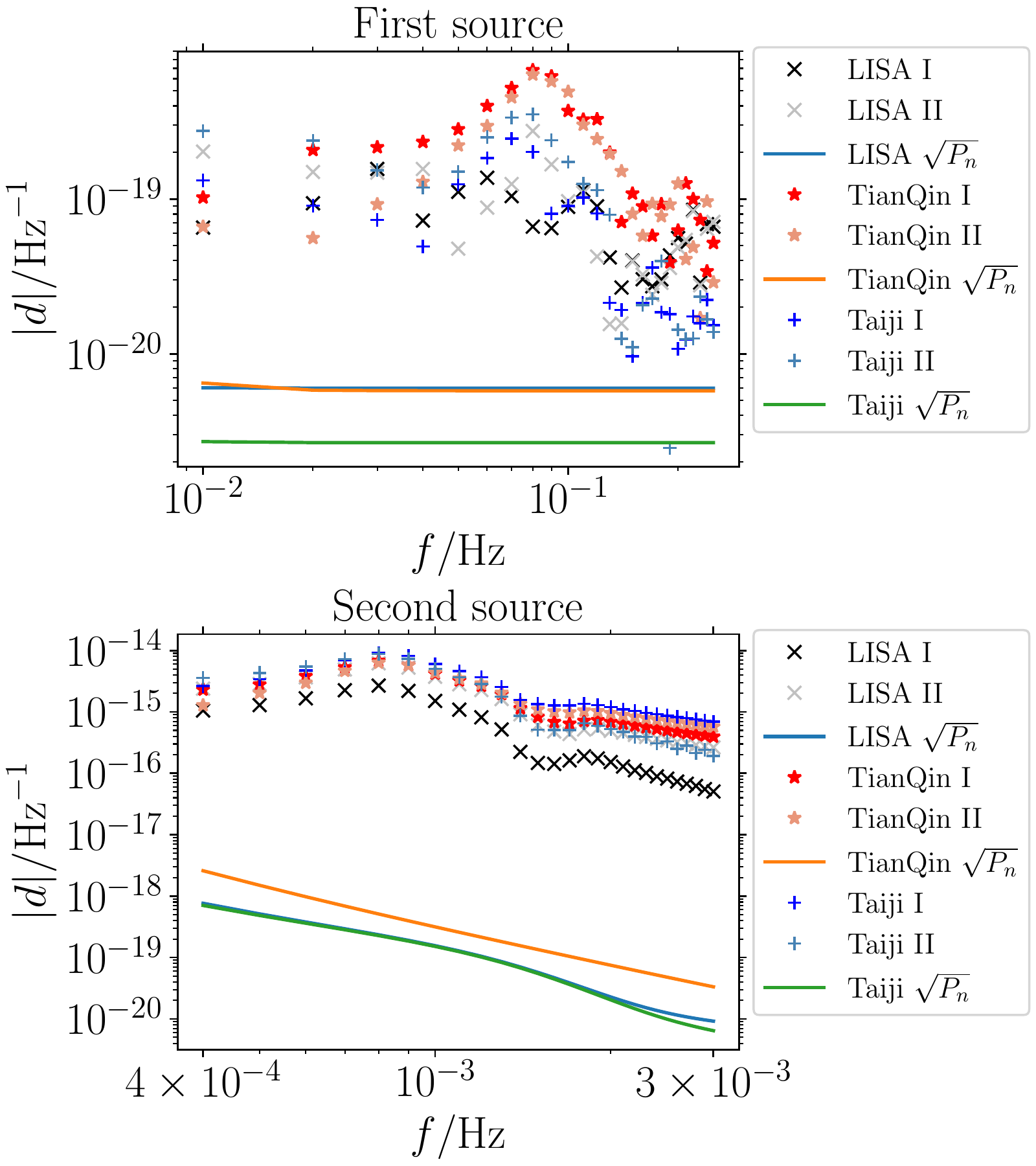}
\caption{The amplitudes of the strain data in LISA, TianQin, and Taiji for the two sources.
The first source has the total mass
$M=10^5M_\odot$.
The sampling frequency and the observation time are set to be 1 Hz and 100 s respectively.
The second source has the total mass $M=10^7M_\odot$.
The sampling frequency and the observation time are set to be 0.1 Hz and 10000 s.
The other parameters of the two sources are $q=2$, $z=1$, $\theta_s=\phi_s=\psi_s=\phi_{\ell m}=\pi/3$ and $\iota=\pi/4$.}
\label{FigData}	
\end{figure}

\begin{figure*}
\includegraphics[width=0.9\textwidth]{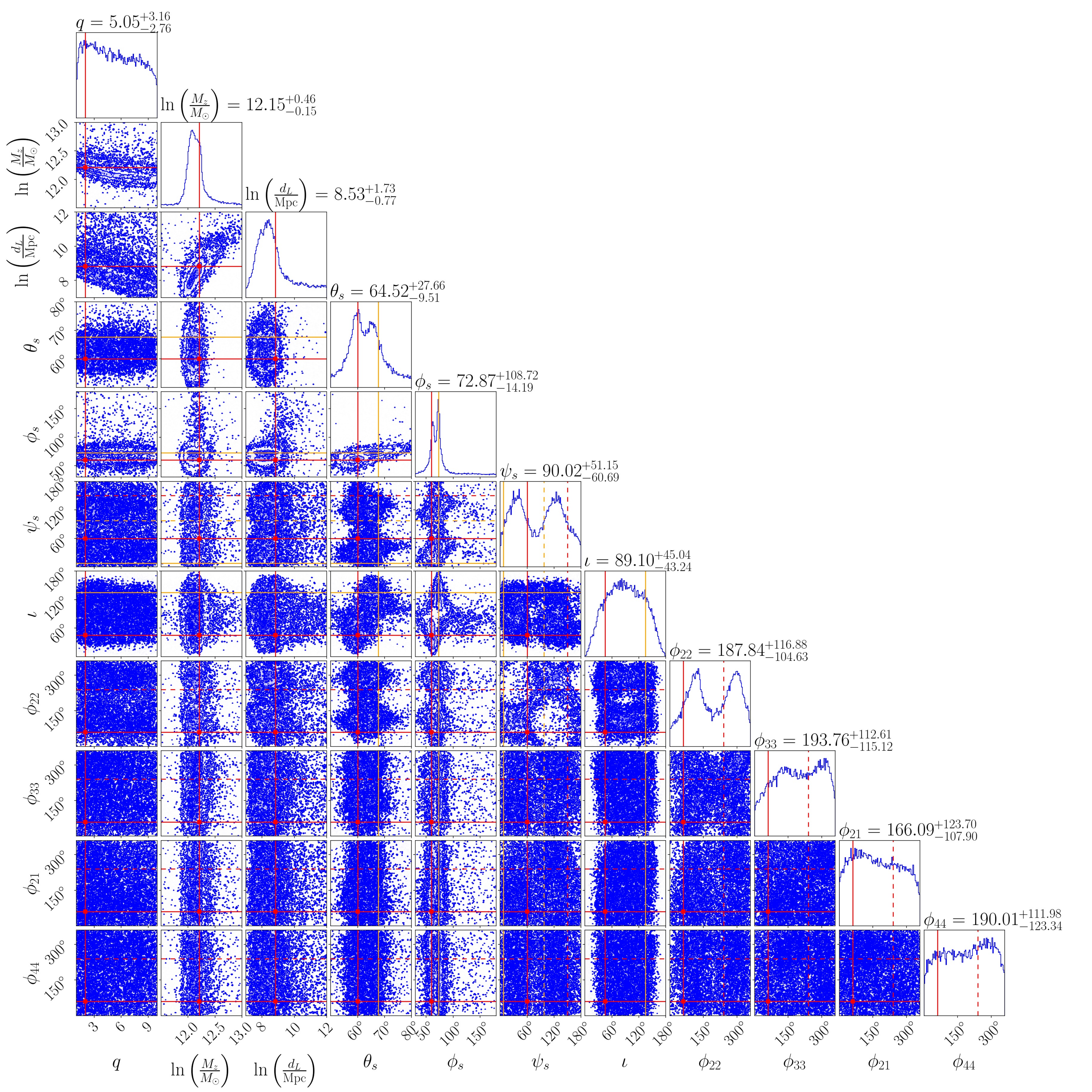}
\caption{The posterior distribution for the first source with LISA.
The red solid lines (dots) represent the true parameters, 
the orange solid lines represent the reflected sky position given by Eq. \eqref{EqDegeneracy1}, 
and the dashed lines represent the degenerate points  given by Eq. \eqref{EqDegeneracy3}. }
\label{FigSample1LISA}	
\end{figure*}

\begin{figure*}
\includegraphics[width=0.9\textwidth]{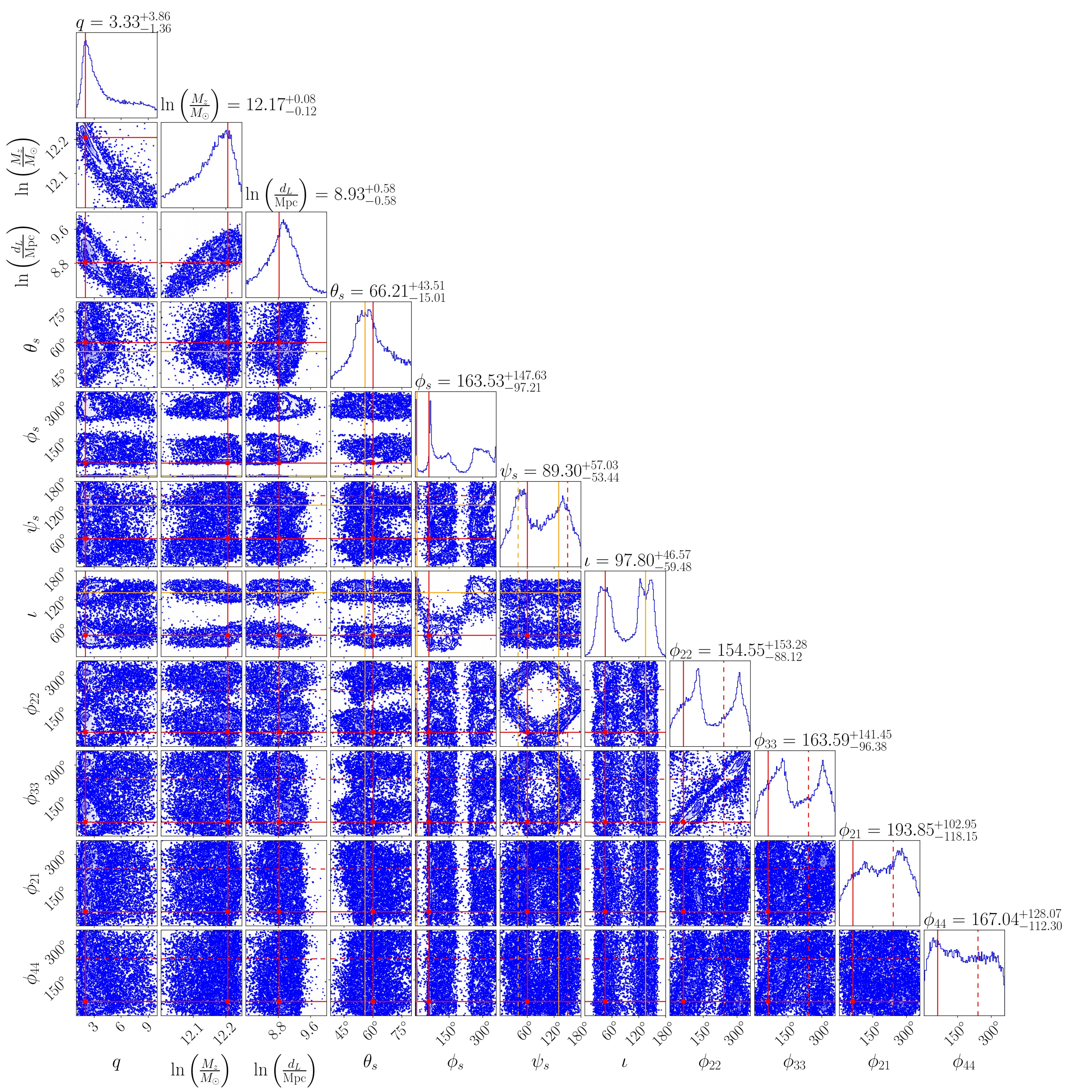}
\caption{The posterior distribution for the first source with TianQin.
The red solid lines (dots) represent the true parameters, 
the orange solid lines represent the reflected sky position given by Eq. \eqref{EqDegeneracy1}, 
and the dashed lines represent the degenerate points given by Eq. \eqref{EqDegeneracy3}. }
\label{FigSample1TQ}
\end{figure*}

\begin{figure*}
\includegraphics[width=0.9\textwidth]{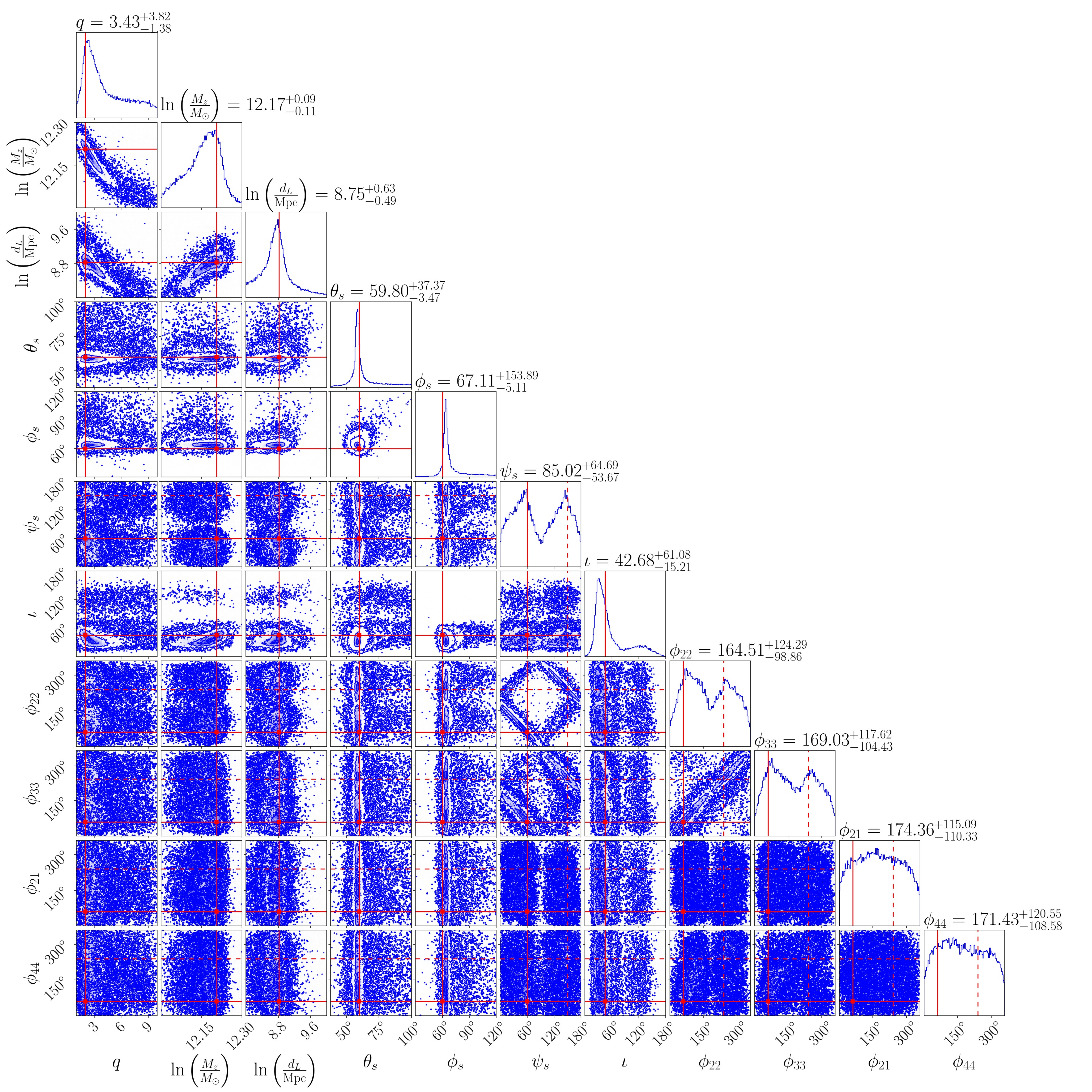}
\caption{The posterior distribution for the first source with the network of LISA and TianQin.
The red solid lines (dots) represent the true parameters, 
the orange solid lines represent the reflected sky position given by Eq. \eqref{EqDegeneracy1}, 
and the dashed lines represent the degenerate points by Eq. \eqref{EqDegeneracy3}.}
\label{FigSample1LISATQ}
\end{figure*}

\begin{figure*}
\includegraphics[width=0.9\textwidth]{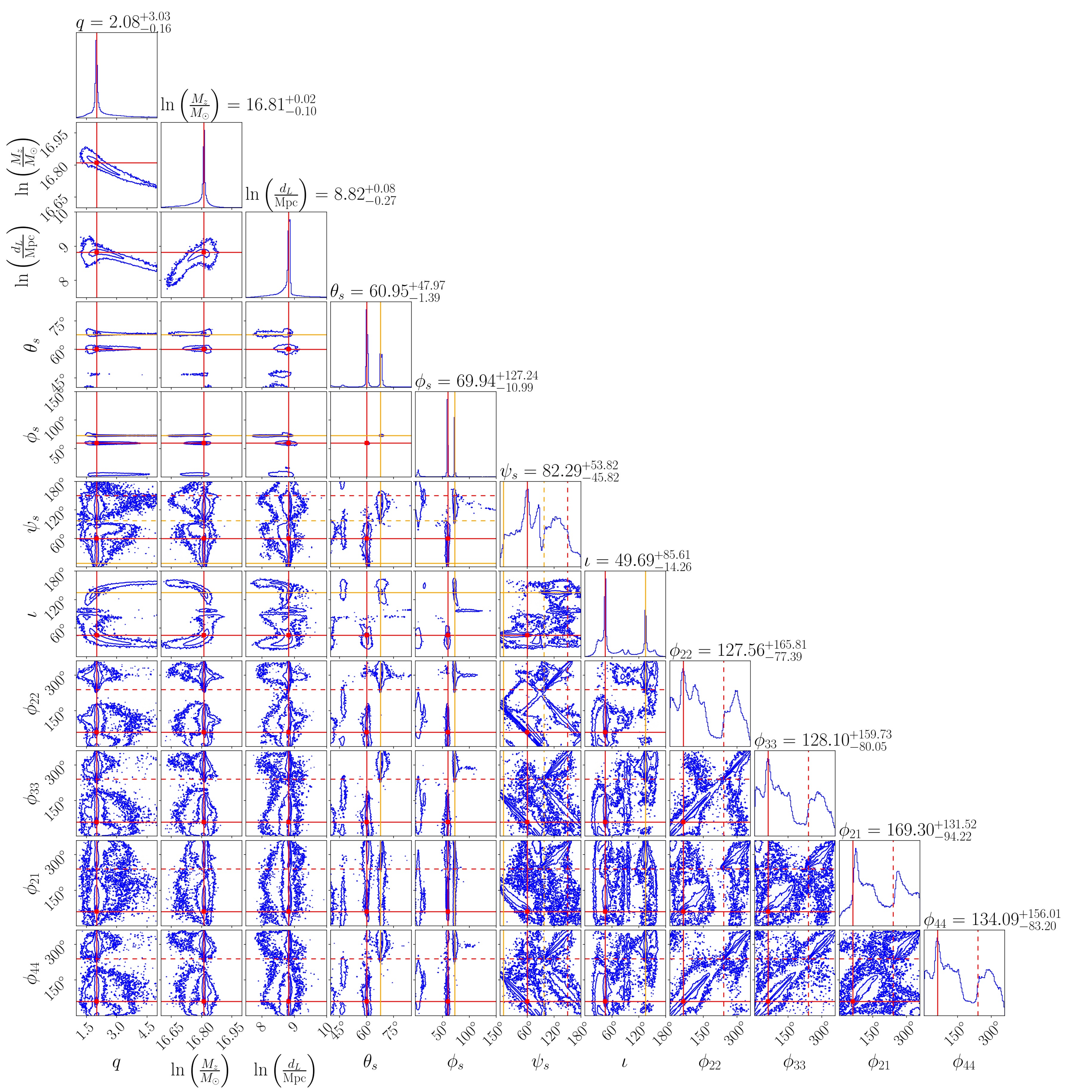}
\caption{The posterior distribution for the second source with LISA.
The red solid lines (dots) represent the true parameters, 
the orange solid lines represent the reflected sky position given by Eq. \eqref{EqDegeneracy1}, 
and the dashed lines represent the degenerate points  given by Eq. \eqref{EqDegeneracy3}.}
\label{FigSample2LISA}	
\end{figure*}

\begin{figure*}
\includegraphics[width=0.9\textwidth]{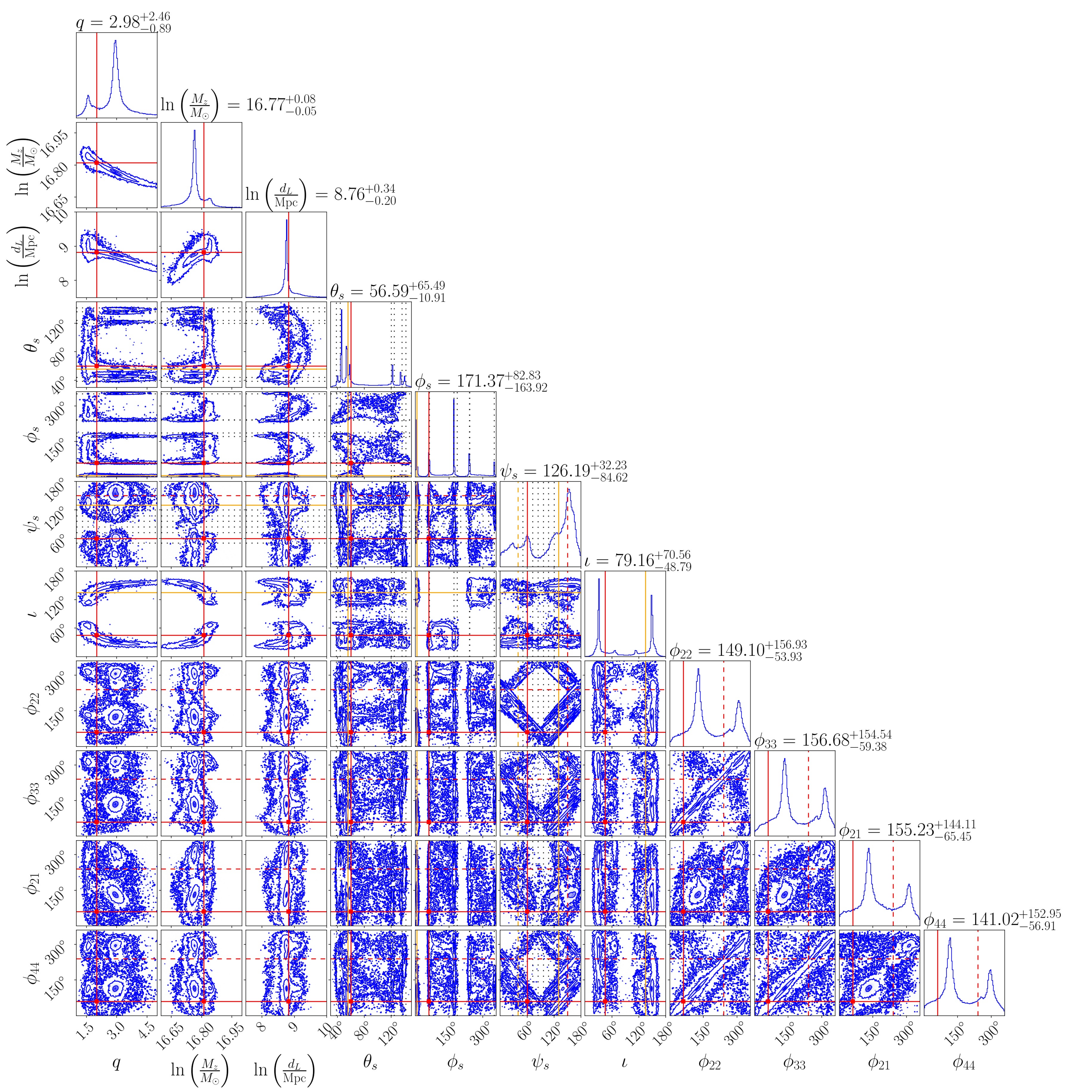}
\caption{The posterior distribution for the second source with TianQin.
The red solid lines (dots) represent the true parameters, 
the orange solid lines represent the reflected sky position given by Eq. \eqref{EqDegeneracy1}, 
the black dotted lines represent the degenerate points given by Eq. \eqref{EqDegeneracy2},
and the dashed lines represent the degenerate points given by Eq. \eqref{EqDegeneracy3}.}
\label{FigSample2TQ}	
\end{figure*}

\begin{figure*}
\includegraphics[width=0.9\textwidth]{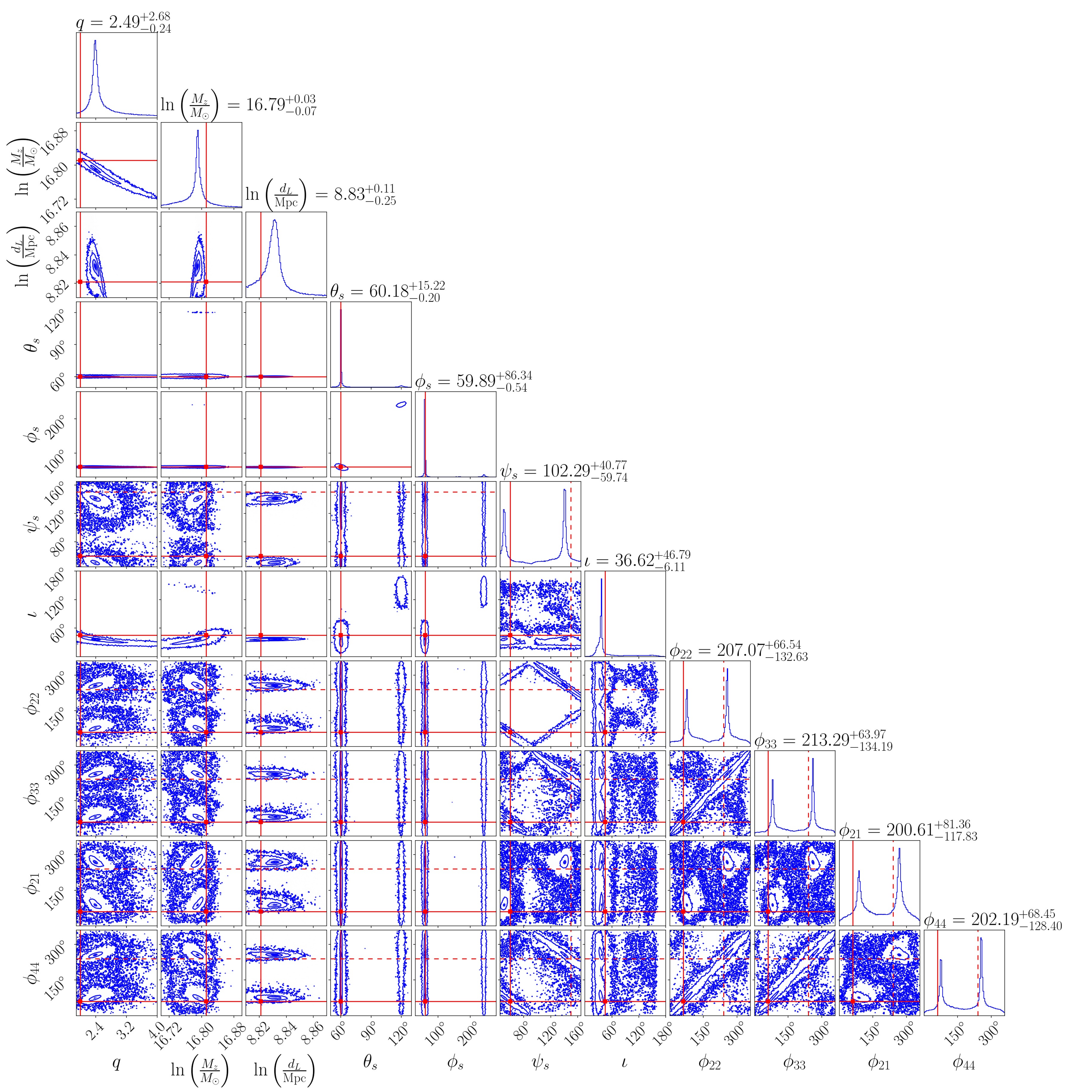}
\caption{The posterior distribution for the second source with the network of LISA and TianQin.
The red solid lines (dots) represent the true parameter values, 
the orange solid lines represent the reflected sky position given by Eq. \eqref{EqDegeneracy1}, and the dashed lines represent the degenerate points  given by Eq. \eqref{EqDegeneracy3}.}
\label{FigSample2LISATQ}
\end{figure*}

The amplitudes of the strain data in LISA, TianQin, and Taiji for the two sources are shown in Fig. \ref{FigData}.
From Fig. \ref{FigData}, we see a high peak which corresponds to the modes $(2,2)$ and $(2, 1)$,
and a low peak behind the high peak which corresponds to the modes $(3,3)$ and $(4, 4)$.
If we work in the heliocentric coordinate, from Eqs. \eqref{EqGWHelioVectors} and \eqref{EqPolar}, we find that the phases of GW signals have the degeneracy
\begin{equation}\label{EqDegeneracy3}
s(\psi_s,\phi_{\ell m})=s(\psi_s\pm\pi)=s(\psi_s\pm\pi/2, \phi_{\ell m}\pm\pi),
\end{equation}
which is not apparent if we work in the detector coordinate.
Thus, all the posterior distributions of $\phi_{\ell m}$ at least have two peaks.
We set the prior distributions of the parameters $\{q, \ln (M/M_\odot), \ln (d_L/\text{Mpc}), \cos \theta_s, \phi_s, \psi_s, \cos\iota, \phi_{\ell m}\}$ to be uniform in the ranges $[1,10]$, $[11.5,20.72]$, $[6.2, 20.34]$, $[-1,1]$, $[0,2\pi]$, $[0,\pi]$, $[-1,1]$, and $[0,2\pi]$, respectively.
The results of Bayesian analysis are shown in Figs. \ref{FigSample1LISA}, \ref{FigSample1TQ}, \ref{FigSample1LISATQ}, \ref{FigSample2LISA}, \ref{FigSample2TQ}, and \ref{FigSample2LISATQ}. The Bayesian results with Taiji are similar to those with LISA, and the results with the 3-network are 
a few times better than those with the network of LISA and TianQin.

For the first source, from Figs. \ref{FigSample1LISA}, \ref{FigSample1TQ}  and \ref{FigSample1LISATQ}, the errors $\{\delta\theta_s,\delta\phi_s\}$ with LISA, TianQin and the network of LISA and TianQin are $\{15^\circ, 34^\circ \}$, $\{25^\circ, 81^\circ\}$ and $\{13^\circ, 27^\circ\}$, respectively, which are a few times worse than the median errors $\{6.4^\circ,4.8^\circ\}$, 
$\{16.8^\circ,20.1^\circ\}$ and 
$\{1.4^\circ,2.4^\circ\}$ with the FIM method given by Figs. \ref{FigLISA}, \ref{FigTQ} and \ref{FigLISATQ}.
For the second source, from Figs. \ref{FigSample2LISA}, \ref{FigSample2TQ}  and \ref{FigSample2LISATQ}, the  errors $\{\delta\theta_s,\delta\phi_s\}$ with LISA, TianQin and the network of LISA and TianQin are $\{6^\circ, 9^\circ\}$, 
$\{15^\circ, 32^\circ\}$ and $\{0.3^\circ, 0.5^\circ\}$, respectively, 
which are about one order worse than the median errors 
$\{0.5^\circ, 0.6^\circ\}$, $\{1.6^\circ,  2.3^\circ\}$ and
$\{0.023^\circ, 0.033^\circ\}$ with the FIM method given by Figs. \ref{FigLISA}, \ref{FigTQ} and \ref{FigLISATQ}.
The FIM results show that the detector network improves the source localization about two order of magnitudes,
while the Bayesian results show that the detector network also plays an important role in eliminating degenerate sky positions. 
In particular, the inherent multimodal distributions for $\theta_s$ and $\phi_s$ with single detector become unimodal distributions with the detector network.
We also find that the estimation errors of $\{q, M_z, d_L, \psi_s, \iota, \phi_{\ell m}\}$ given by the two methods are consistent with each other within one order of magnitude.

\section{conclusion}
\label{SecConclusion}

We derive the analytical formulas of the frequency-domain ringdown signals with the harmonic phases, 
the rotation period of the geocentric detector, and the detector's arm length.
The analytical formulas help a lot to reduce the computation time in the FIM analysis.
We show the median errors of the parameter estimation and source localization with ringdown signals from binaries with different total masses and different redshifts.
We find that for binaries with the total mass $M\ge 10^5\ M_\odot$, space-based GW detectors can effectively estimate parameters and localize sources with the ringdown signal.
For the ringdown frequencies $0.5f^*\le f \le 5f^*$, we find that the transfer function dramatically improves the parameter estimation and source localization.
Thus the transfer function of LISA, Taiji, and TianQin can improve their source localization for binaries with the redshifted total masses $1.7\times10^5\ M_\odot\le M_z \le 1.7\times10^6\ M_\odot$, $2\times10^5\ M_\odot\le M_z \le 2\times10^6\ M_\odot$, and $1.2\times10^4\ M_\odot\le M_z \le 1.2\times10^5\ M_\odot$, respectively.
We also find that for BH binaries at the same distance,
LISA has larger SNR and better parameter estimation and source localization for the BH binary with $M_z=6.5\times10^6\ M_\odot$,
TianQin has larger SNR and better parameter estimation and source localization for the BH binary with $M_z=3\times10^6\ M_\odot$,
and LISA-TianQin network has larger SNR and better parameter estimation and source localization for the BH binary with $M_z=4\times10^6\ M_\odot$.

As for the dependence of the parameter estimation and source localization on the sky position, we find that the detector has the best angular resolution for sources along the detector plane, where the relative difference of source location reaches the maximum, but it has the worst angular resolution for sources perpendicular to the detector plane, where the relative difference of source location is the minimum.
Except $\{\theta_d,\phi_d\}$, the worst parameter estimation accuracy  occurs for sources along the detector plane, where the tensor response is the minimum, 
and the best parameter estimation accuracy occurs for sources along the two poles, where the tensor response is the maximum.
However, the estimation error of the polarization angle $\psi_d$ becomes the worst for sources along the two poles, because of its coupling with $\phi_d$. 
In fact, the difference of the parameter estimation for sources at different locations is within one order of magnitude in most cases.

To check the FIM results, 
we use Bayesian inference method to analyze two typical binaries. 
We find that the results of the parameter estimation and source localization given by the two methods are consistent with each other.
Thus we expect that in real data analysis, 
especially in the case of GW detector network, 
the results of the parameter estimation and source localization given by Bayesian analysis are close to the median errors given by Figs. \ref{FigLISA}, \ref{FigTQ} and \ref{FigLISATQ} within one order of magnitude in most cases.

The network of space-based GW detectors not only improves the sky localization accuracy by two or even three orders of magnitude  compared with individual detector,
but also avoids the reflected sky position,
and it is sensitive to GWs from almost all directions.
Thus, to reach the scientific goals efficiently for GW observations,
the combined network is extremely important for not only ground-based GW detectors, but also space-based GW detectors. 
We provide a useful approach to understanding parameter estimation with ringdown signals in space-based GW detectors, which is important to quickly understand the range of possibilities in these detectors.
The results are helpful for exploring the scientific potential of space-based GW detectors.

\begin{acknowledgments}
Chunyu Zhang thanks Vishal Baibhav and Vitor Cardoso for useful discussions.
This research is supported in part by the National Key Research and Development Program of China under Grant No. 2020YFC2201504,
the National Natural Science
Foundation of China under Grant No. 11875136,
and the Major Program of the National Natural Science Foundation of China under Grant No. 11690021.
\end{acknowledgments}

\appendix

\begin{widetext}

\section{Analytical formulas}
\label{AppAnalytical}

In this paper, we use the following formulas:
\begin{equation}\label{EqIa}
\begin{split}
&I_a(\omega_{\ell m},\tau_{\ell m},\phi_{\ell m})=\int_{0}^{+\infty}e^{-\frac{t}{\tau_{\ell m}}}\cos(\omega_{\ell m} t-\phi_{\ell m})e^{-i\omega t}dt
=\frac{I_{a1}\cos\phi_{\ell m}-I_{a2}\tau_{\ell m}\omega_{\ell m}\sin\phi_{\ell m}}{I_{a3}}\tau_{\ell m},
\\
&I_{a1}=(1+i\omega\tau_{\ell m})\left[1+2i\omega\tau_{\ell m}-\tau^2_{\ell m}\left(\omega^2-\omega^2_{\ell m}\right)\right],
\\
&I_{a2}=\tau^2_{\ell m}\left(\omega^2-\omega_{\ell m}^2\right)-1-2i\omega\tau_{\ell m},
\\
&I_{a3}=\left[\tau_{\ell m}\left(\omega-\omega_{\ell m}\right)-i\right]^2\left[\tau_{\ell m}\left(\omega+\omega_{\ell m}\right)-i\right]^2,
\end{split}
\end{equation}

\begin{equation}\label{EqIb}
\begin{split}
&I_b(\phi_d,\omega_{\ell m},\tau_{\ell m},\phi_{\ell m})=\int_{0}^{+\infty}\sin\left[2\left(\phi_d-\omega_{tq}t\right)\right]e^{-\frac{t}{\tau_{\ell m}}}\cos(\omega_{\ell m} t-\phi_{\ell m})e^{-i\omega t}dt
\\
&\qquad\qquad\qquad\qquad\quad=\frac{I_{b1}\cos\phi_{\ell m}-I_{b2}\tau_{\ell m}\omega_{\ell m}\sin\phi_{\ell m}}{I_{b3}}\tau_{\ell m},
\\
&I_{b1}=2\tau_{\ell m}\omega_{tq}\left[\tau^2_{\ell m}\left(\omega^2+\omega_{\ell m}^2-4\omega_{tq}^2\right)-2i\omega\tau_{\ell m}-1\right]\cos2\phi_d
\\
&\qquad-(1+i\omega\tau_{\ell m})\left[\tau^2_{\ell m}\left(\omega^2-\omega^2_{\ell m} -4\omega_{tq}^2\right)-2i\omega\tau_{\ell m}-1\right]\sin(2\phi_d),
\\
&I_{b2}=\left[\tau^2_{\ell m}\left(\omega^2-\omega_{\ell m}^2+4\omega_{tq}^2\right)-2i\omega\tau_{\ell m}-1\right]\sin(2\phi_d)+4\tau_{\ell m}\omega_{tq}(1+i\omega\tau_{\ell m})\cos2\phi_d,
\\
&I_{b3}=\left[\tau_{\ell m}\left(\omega-\omega_{\ell m}-2\omega_{tq}\right)-i\right]
\left[\tau_{\ell m}\left(\omega+\omega_{\ell m}-2\omega_{tq}\right)-i\right]
\\
&\qquad\times\left[\tau_{\ell m}\left(\omega-\omega_{\ell m}+2\omega_{tq}\right)-i\right]
\left[\tau_{\ell m}\left(\omega+\omega_{\ell m}+2\omega_{tq}\right)-i\right],
\end{split}
\end{equation}

\begin{equation}
\begin{split}
&\int_{0}^{+\infty}-e^{-\frac{t}{\tau_{\ell m}}}\sin(\omega_{\ell m} t-\phi_{\ell m})e^{-i\omega t}dt
=I_a(\omega_{\ell m},\tau_{\ell m},\phi_{\ell m}-\frac{\pi}{2}),\\
&\int_{0}^{+\infty}\cos\left[2\left(\phi_d-\omega_{tq}t\right)\right]e^{-\frac{t}{\tau_{\ell m}}}\cos(\omega_{\ell m} t-\phi_{\ell m})e^{-i\omega t}dt
=I_b(\phi_d+\frac{\pi}{4},\omega_{\ell m},\tau_{\ell m},\phi_{\ell m}),
\\
&\int_{0}^{+\infty}-\sin\left[2\left(\phi_d-\omega_{tq}t\right)\right]e^{-\frac{t}{\tau_{\ell m}}}\sin(\omega_{\ell m} t-\phi_{\ell 	m})e^{-i\omega t}dt=I_b(\phi_d,\omega_{\ell m},\tau_{\ell m},\phi_{\ell m}-\frac{\pi}{2}),
\\
&\int_{0}^{+\infty}-\cos\left[2\left(\phi_d-\omega_{tq}t\right)\right]e^{-\frac{t}{\tau_{\ell m}}}\sin(\omega_{\ell m} t-\phi_{\ell 	m})e^{-i\omega t}dt=I_b(\phi_d+\frac{\pi}{4},\omega_{\ell m},\tau_{\ell m},\phi_{\ell m}-\frac{\pi}{2}),
\\
&\int_0^L\left(e^{2\pi if[-2L+(1-\mu)\lambda]}+e^{2\pi if[-L+\lambda-\mu(L-\lambda)]}\right)d\lambda =2L{\cal T}(f,\mu).
\end{split}
\end{equation}
Here $\omega=2\pi f$.

\section{The coordinate transformation}
\label{CoorTrans}

In this section, we give the coordinate transformation formulas  from the heliocentric coordinate $\{\hat{i},\hat{j},\hat{k}\}$ to the detector coordinate $\{\hat{x},\hat{y},\hat{z}\}$.
The Euler rotation matrices are
\begin{equation}\label{EqEuler}
R_x(\theta)=
\begin{bmatrix}
1 & 0 & 0 \\
0 & \cos\theta & -\sin\theta \\
0 & \sin\theta & \cos\theta
\end{bmatrix},
R_y(\theta)=
\begin{bmatrix}
\cos\theta  & 0 & \sin\theta \\
0           & 1 & 0 \\
-\sin\theta & 0 & \cos\theta
\end{bmatrix},
R_z(\theta)=
\begin{bmatrix}
\cos\theta  &  -\sin\theta & 0\\
\sin\theta  &  \cos\theta  & 0\\
0           &  0           & 1
\end{bmatrix}.
\end{equation}
GW coordinate basis vectors in the heliocentric coordinate are given by
\begin{equation}
\begin{split}
\{\hat{m},\hat{n},\hat{o}\}=&\{\hat{i},\hat{j},\hat{k}\}\times R_z\left(\phi_{s}-\pi\right) R_y\left(\pi-\theta_{s}\right) R_z\left(\psi_s\right)\\
=&\begin{bmatrix}
\cos\theta_s\cos\phi_s\cos\psi_s+\sin\phi_s\sin\psi_s  &  \sin\phi_s\cos\psi_s-\cos\theta_s\cos\phi_s\sin\psi_s & -\sin\theta_s\cos\phi_s
\\
\cos\theta_s\sin\phi_s\cos\psi_s-\cos\phi_s\sin\psi_s  &  -\cos\phi_s\cos\psi_s-\cos\theta_s\sin\phi_s\sin\psi_s  & -\sin\theta_s\sin\phi_s
\\
-\sin\theta_s\cos\psi_s  &  \sin\theta_s\sin\psi_s   & -\cos\theta_s
\end{bmatrix},
\end{split}
\end{equation}
where $(\theta_s,\phi_s)$ are the source location, and $\psi_s$ is the polarization angle.

For TianQin, the detector coordinate basis vectors are
\begin{equation}\label{tqcoor}
\{\hat{x},\hat{y},\hat{z}\}=\{\hat{i},\hat{j},\hat{k}\}\times R_z\left(\phi_{tq}-\frac{\pi}{2}\right) R_x\left(-\theta_{tq}\right)\,,
\end{equation}	
where $(\theta_{tq},\phi_{tq})=(94.7^\circ,120.5^\circ)$, and $\omega_{tq}=2\pi/T_{tq}=1.99\times10^{-5}$ Hz is the rotation frequency of TianQin.
For LISA, the detector coordinate basis vectors are
\begin{equation}\label{lisacoor}
\{\hat{x},\hat{y},\hat{z}\}=\{\hat{i},\hat{j},\hat{k}\}\times R_z\left(\omega_{e}t\right) R_y\left(-\frac{\pi}{3}\right) R_z\left(-\omega_{lisa}t\right)\,,
\end{equation}	
where $\omega_{e}=\omega_{lisa}=1.99\times10^{-7}$ Hz is the rotation frequency of the Earth and LISA around the Sun.
Thus the source parameters in the detector coordinate are
	
\begin{equation}\label{DetParams}
\begin{split}
\theta_d=\arccos(-\hat{o}\cdot\hat{z}),
\phi_d=2\arctan\left(\frac{-\hat{o}\cdot\hat{y}}{\sin\theta_d-\hat{o}\cdot\hat{x}}\right),
\psi_d=2\arctan\left(\frac{\hat{n}\cdot\hat{z}}{\sin\theta_d-\hat{m}\cdot\hat{z}}\right).
\end{split}
\end{equation}

\section{Simulated sources}
\label{AppSource}
For each binary with the same total mass and redshift,
we use Monte Carlo simulation to generate 1000 sources.
In this Appendix, we show the distribution of the parameters for the simulated sources in Fig. \ref{FigSources}.

\begin{figure*}
\includegraphics[width=0.9\textwidth]{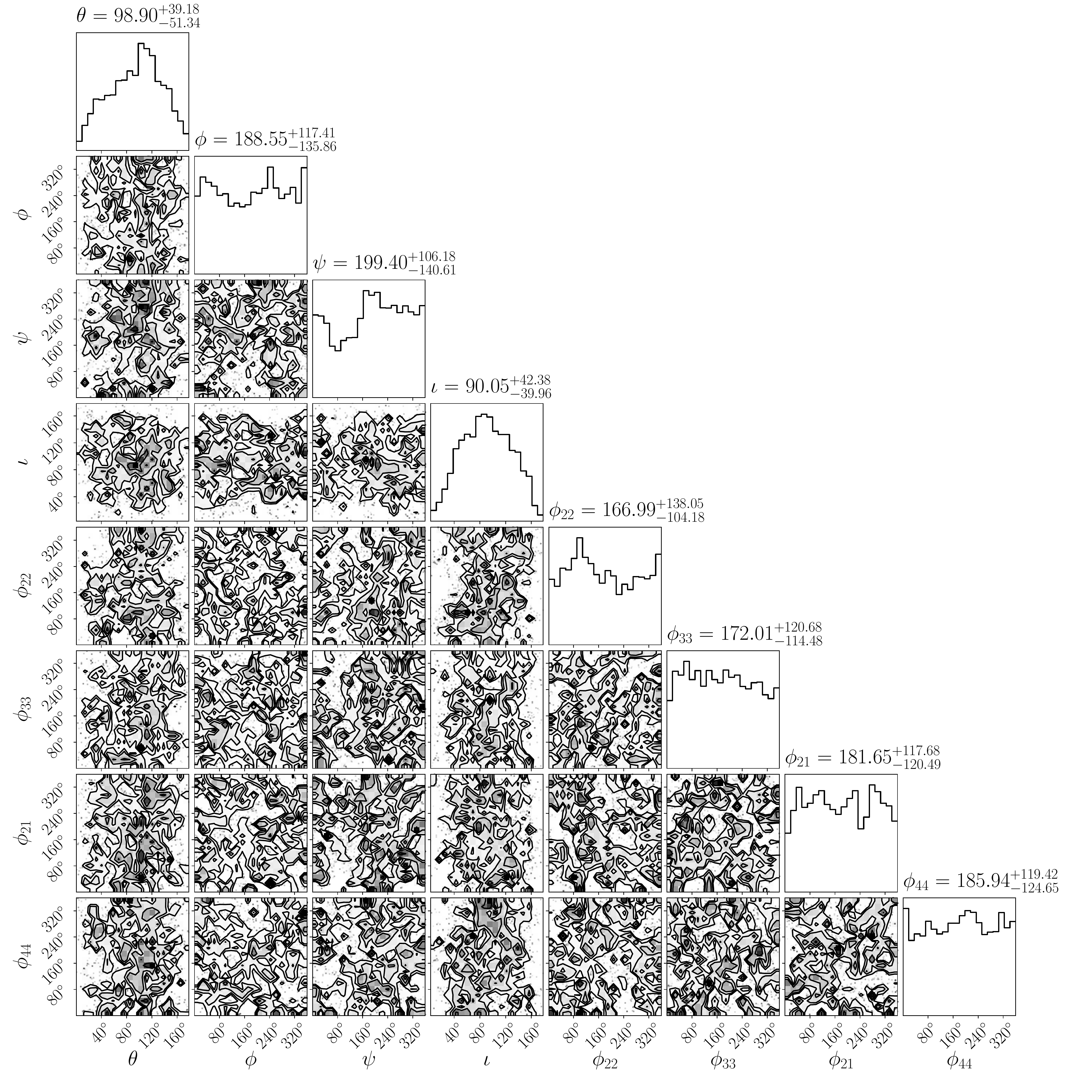}
\caption{The parameter distribution of simulated sources.}
\label{FigSources}	
\end{figure*}


\end{widetext}



%

\end{document}